\documentclass[traditabstract,referee]{aa}
\usepackage{natbib}
\usepackage{graphicx}
\usepackage{epsfig}
\usepackage{txfonts}

\def\ms{\,m\,s$^{-1}$}         
\def\msol{M$_\odot$}             
\def\mjup{M$_{Jup}$}             
\def\a{HD\,103720\,}
\def\b{HD\,564\,}

\def\d{HD\,108341\,}
\def\e{HD\,113538\,}
\def\f{BD\,-114672\,}
\begin{document}

\title{ The HARPS search for southern extra-solar planets\thanks{Based on observations made with the HARPS instrument on the
ESO 3.6 m telescope at La Silla (Chile), under the GTO program ID 072.C-0488, 183.C-0972 and the regular programs: 085.C-0019, 087.C-0831,
089.C-0732, 090.C-0421, 091.C-0034, and 092.C-0721}
}

\subtitle{XXXVI: Five new long-period giant planets and a system update}

\author{
C. Moutou\inst{1,2}
\and G. Lo Curto\inst{3}
\and M. Mayor\inst{4}
\and F. Bouchy\inst{2}
\and W. Benz\inst{5}
\and C. Lovis\inst{4}
\and D. Naef\inst{4}
\and F. Pepe\inst{4}
\and D. Queloz\inst{4}
\and N.C. Santos\inst{6,7}
\and D. S\'egransan\inst{4}
\and S. G. Sousa\inst{6,7}
\and S. Udry\inst{4}
}

\offprints{C. Moutou}

\institute{
Canada France Hawaii Telescope Corporation, Kamuela 96743, USA, {\email moutou@cfht.hawaii.edu}
\and 
Aix Marseille Universit\'e, CNRS, LAM (Laboratoire d'Astrophysique de Marseille) UMR 7326, Marseille, France
\and 
ESO, Alonso de C\'ordova 3107, Vitacura, Casilla 19001, Santiago de Chile, Chile
\and
Observatoire de Gen\`eve, Universit\'e de Gen\`eve, 51 chemin des Maillettes, 1290 Sauverny, Switzerland.
\and
Physikalisches Institut Universit\"at Bern, Sidlerstrasse 5, 3012
Bern, Switzerland
\and
Instituto de Astrof\'{i}sica e Ci\^encias do Espaco, Universidade do Porto, CAUP, Rua das Estrelas, 4150-762 Porto, Portugal
\and
Centro de Astrof\'{i}sica e Departamento de F\'{i}sica e Astronomia, Faculdade de Ci\^encias, Universidade do Porto, Rua das Estrelas, 4150-762 Porto, Portugal}

\date{Received ; accepted }

\abstract{We describe radial-velocity time series obtained by HARPS on the 3.60 m telescope in La Silla (ESO, Chile) over ten years and report the discovery of five new giant exoplanets in distant orbits; these new planets orbit the stars HD 564, HD 30669, HD 108341, and BD-114672. Their periods range from 492 to 1684 days, semi-major axes range from 1.2 to 2.69 AU, and eccentricities range from 0 to 0.85. Their minimum mass ranges from 0.33 to 3.5 \mjup. We also refine the parameters of two planets announced previously around HD 113538, based on a longer series of measurements. The planets have a period of 663$\pm$8 and 1818$\pm$25 days, orbital eccentricities of 0.14$\pm$0.08 and 0.20$\pm$0.04, and minimum masses of 0.36$\pm$0.04 and 0.93$\pm$0.06 \mjup. Finally, we report the discovery of a new hot-Jupiter planet around an active star, HD 103720; the planet has a period of 4.5557$\pm$0.0001 days and a minimum mass of 0.62$\pm$0.025 \mjup. We discuss the fundamental parameters of these systems and limitations due to stellar activity in quiet stars with typical 2\ms\ radial velocity precision.}


\maketitle

\section{Introduction}

\begin{figure*}[h]
\epsfig{file=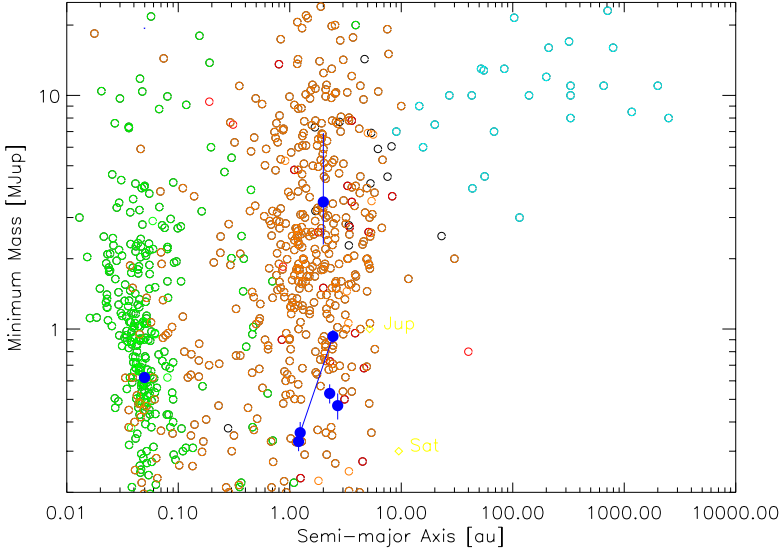,width=0.9\textwidth}
\caption{Semi-major axis versus projected mass for planets farther than 0.01AU and more massive than 0.2 \mjup. The symbol color indicates the discovery method (blue: direct imaging; orange: RV; green: transits; black: microlenses).}
\label{smam}
\end{figure*}

Thanks to radial-velocity (RV) surveys collecting data since more than 20 years, on one hand, and advanced technologies in high-contrast imaging, on the other hand, the parameter spaces of exoplanets discovered by these two methods are now joined in the regime of massive planets (Fig. \ref{smam}). Both exoplanet populations rejoin at a semi-major axis of 8-10 AU (i.e., at Saturn's orbit). The types of stars around which these exoplanets are discovered, however, differ much as a result of the detection bias of these two methods: quiet, FGKM stars are more favorable to RV survey stars, and very young (hence, active) nearby stars to direct imaging surveys, these last ones being mostly of AF spectral types. This complementarity is important for obtaining a global architecture of planetary systems and constrain a planet formation scenario, and some overlap is essential to calibrate measured quantities, especially the mass, since it is inferred from photometric measurements in the case of imaged systems \citep[e.g.,][]{spiegel12,allard13}.
With new upcoming instruments such as GPI \citep{macintosh}  and SPHERE \citep{beuzit}, more giant planets in the 5-10 AU range will be discovered, some of which will hopefully have a corresponding signature in the RV parameter space, which will
allow a gravitational estimate on the planet mass in addition to the spectrophotometric characterization from direct imaging.

In a similar way to how speckles limit the contrast at which a planet can be directly imaged, the search for long-period exoplanets in RV surveys suffers from another noise due to the host star, with activity at the stellar surface generating a RV jitter \citep[e.g.,][]{saar98,wright05,boisse11,dumusque11}. With the amplitude of the Doppler wobble decreasing with the orbital distance, even giant planets have signatures whose amplitude can be buried in stellar noise, given that activity-driven RV noise has multiple frequency components, including the one due to several-year long magnetic cycles \citep{santos10}. To correct for stellar activity, very high precision on the usual diagnostics parameters is needed: FWHM and bisector span of the stellar cross-correlation function, and the CaII index.

In this paper, we present four new giant planets discovered by the RV spectrograph HARPS in the volume-limited survey \citep{locurto10} and discuss their properties: the hot-Jupiter HD 103720 b, and long-period giant planets around HD 564, HD 30669, and HD 108341. In addition, we update the analysis of two systems: HD 113538, composed of two planets for which the best-fit parameters are refined, and BD -114672, whose signal was previously identified to be due to a long-term magnetic cycle \citep{moutou11}. Host stars are described in Sect. \ref{star}, radial-velocity observations and analyses in Sect. \ref{planet}, and Sect. \ref{ccl} presents a discussion of these new results.


\section{Characteristics of the host stars}
\label{star}

Because they are part of the volume-limited HARPS survey, all stars presented in this analysis are less than 57.5 pc away from the Sun. This particular sample is composed of six medium-age G and K stars.

The spectroscopic analysis was performed on the combined HARPS spectra by measuring a set of FeI and FeII weak lines with the
code ARES \citep{sousa07} and was then used in an homogeneous spectroscopic analysis \citep[e.g.,][]{sousa08}. For all stars except \e and \f, the stellar parameters ($T_{\mathrm{eff}}$, log $g,$ and $\mathrm{[Fe/H]}$) were taken from the catalog of \citet{sousa11}. The Hipparcos parallaxes of \citet{vanleeuwen07} were used for the luminosity estimation, and finally, mass, radius, and age of the stars were derived from comparisons with theoretical tracks with a Bayesian analysis\footnote{http://stev.oapd.inaf.it/cgi-bin/param} \citep{dasilva06}. All parameters are listed in Table \ref{stars}. For \e and \f, updates are provided for the spectroscopic analysis, and in addition, masses (and radii and ages) have been corrected from the erroneous values given in \citet{moutou11}; these incorrect mass values were due to missing information in the evolutionary track interface for low-mass stars, which are solved now \citep{chen14}.

Activity indicators ($S_{MW}$ and $\log R'_{\mathrm{HK}}$) were individually estimated on each spectrum, and their range of values is given in Table \ref{stars}. Rotation periods of the stars were estimated from their correlations with the chromospheric activity, as described in \citet{noyes84,mamajek08}. HD 30669 and HD 108341 were less affected by chromospheric activity during these past ten years. 

\begin{figure*}[h]
\epsfig{file=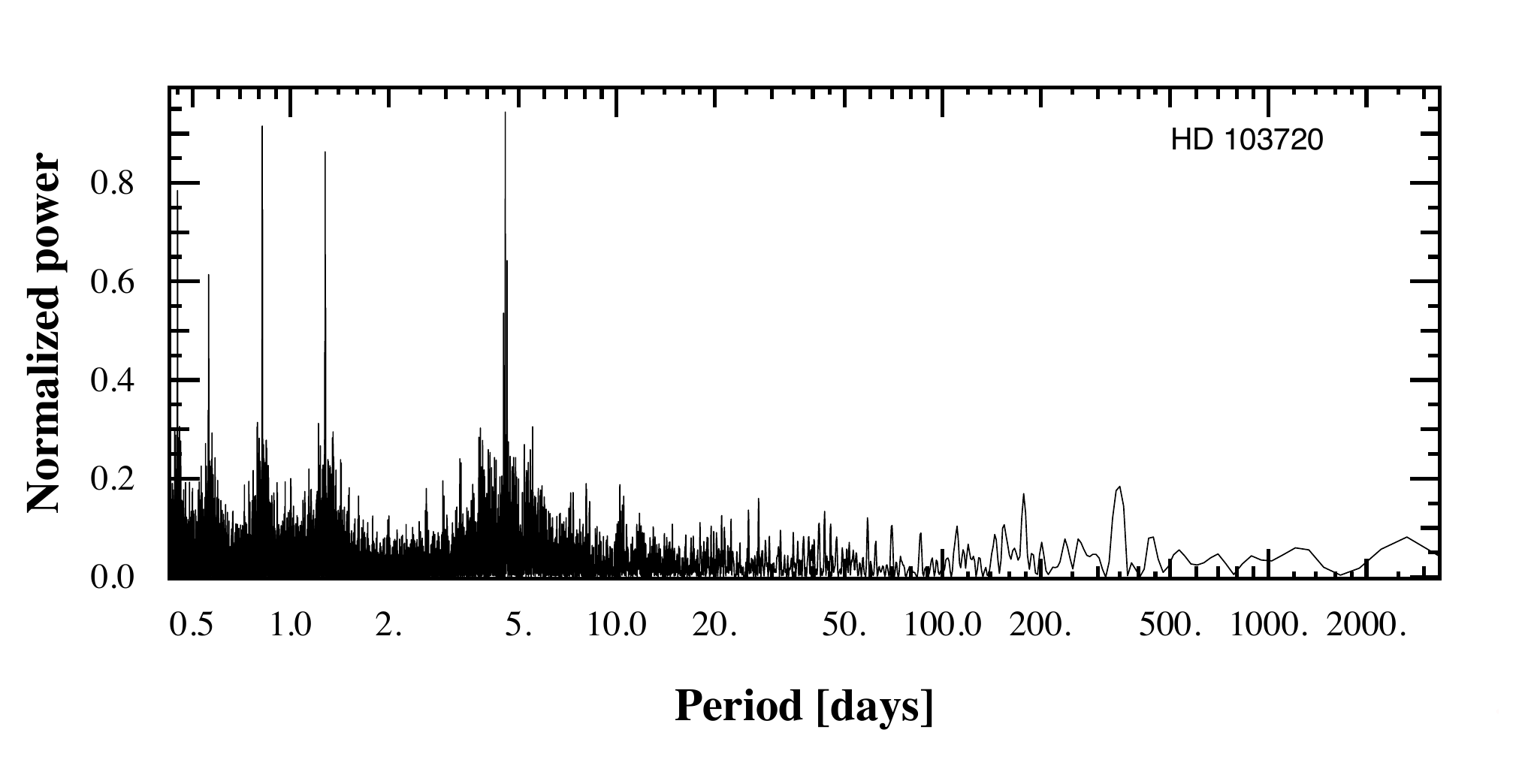,width=0.48\textwidth}\epsfig{file=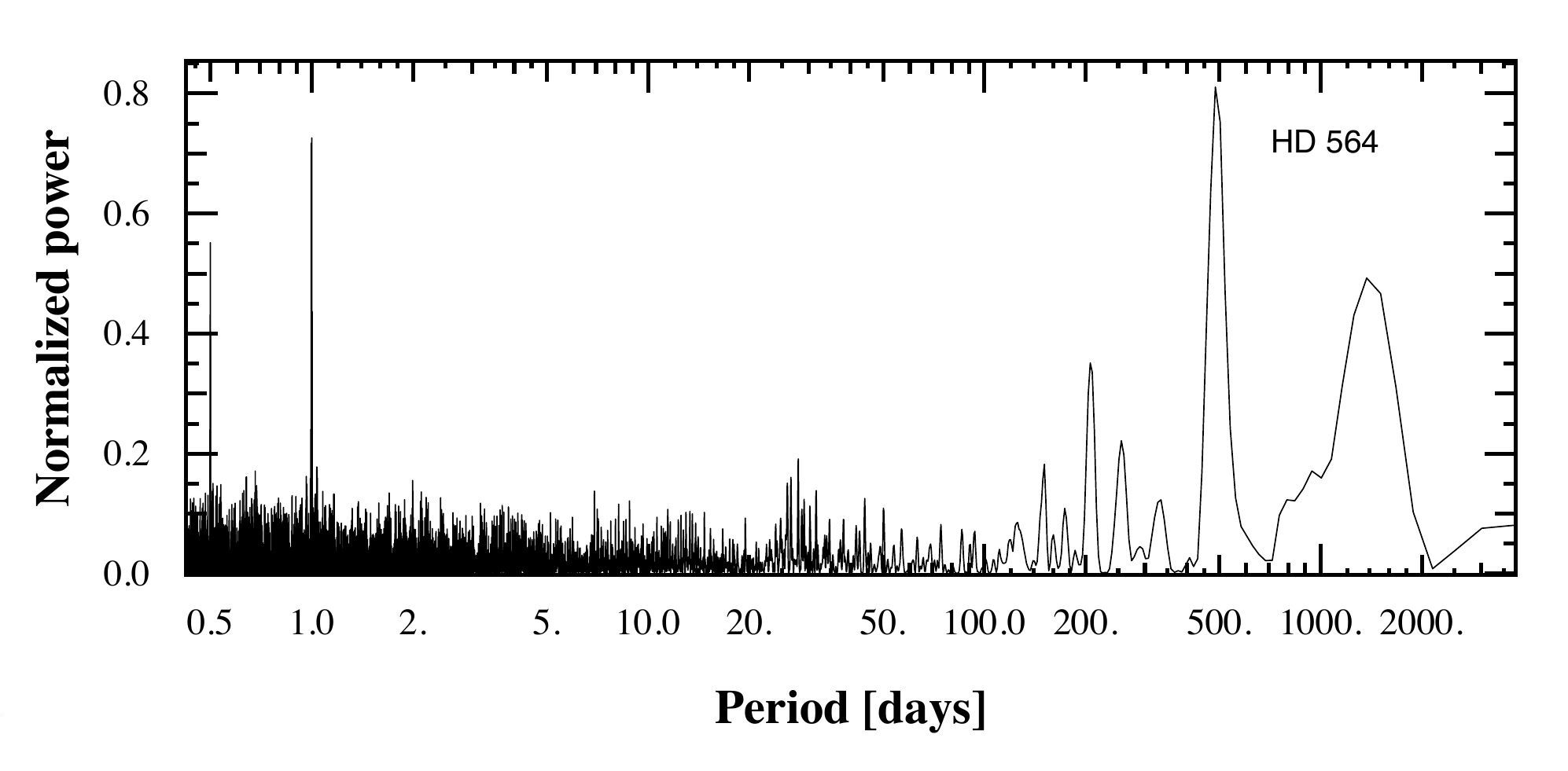,width=0.48\textwidth}
\epsfig{file=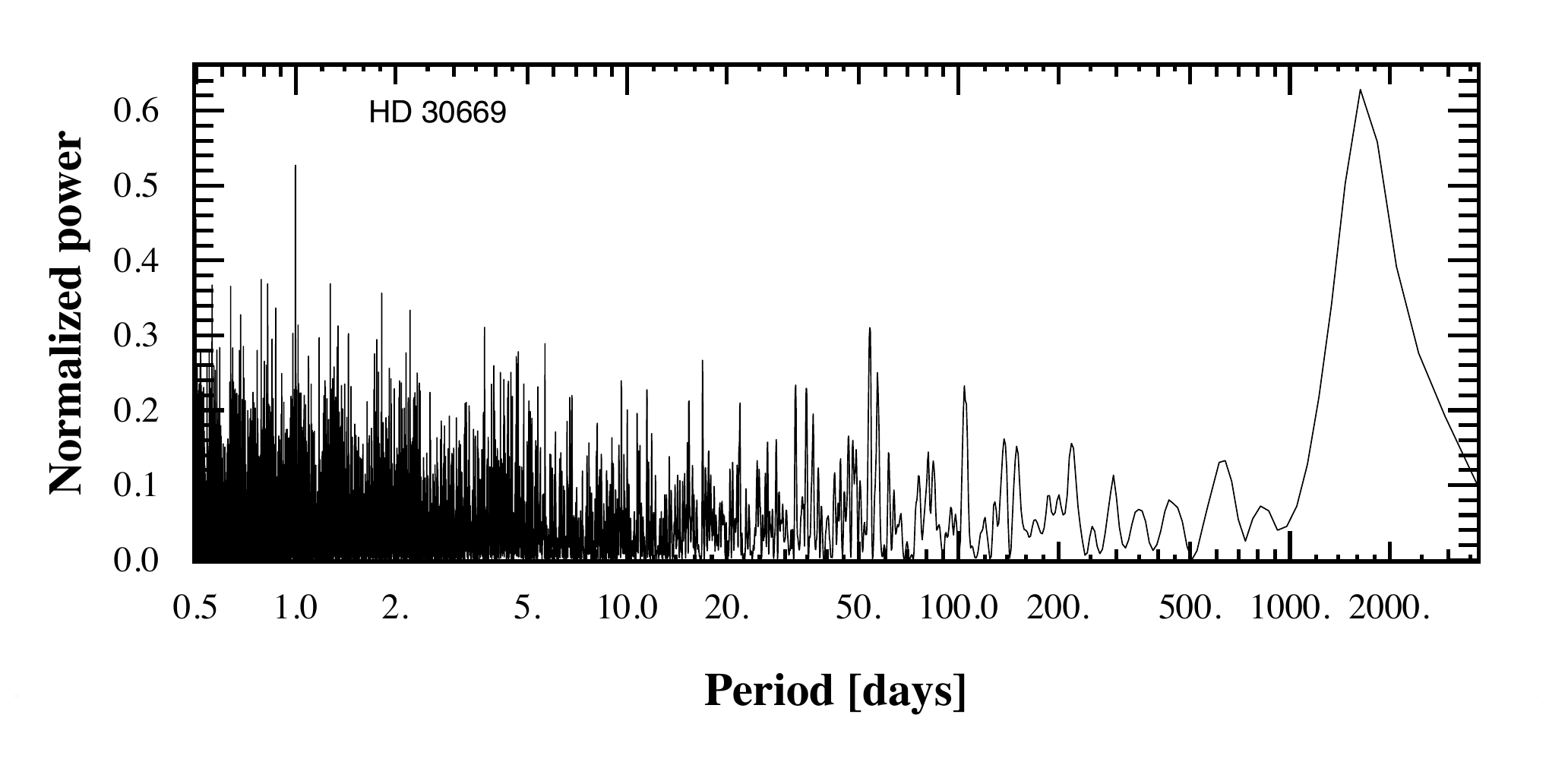,width=0.48\textwidth}\epsfig{file=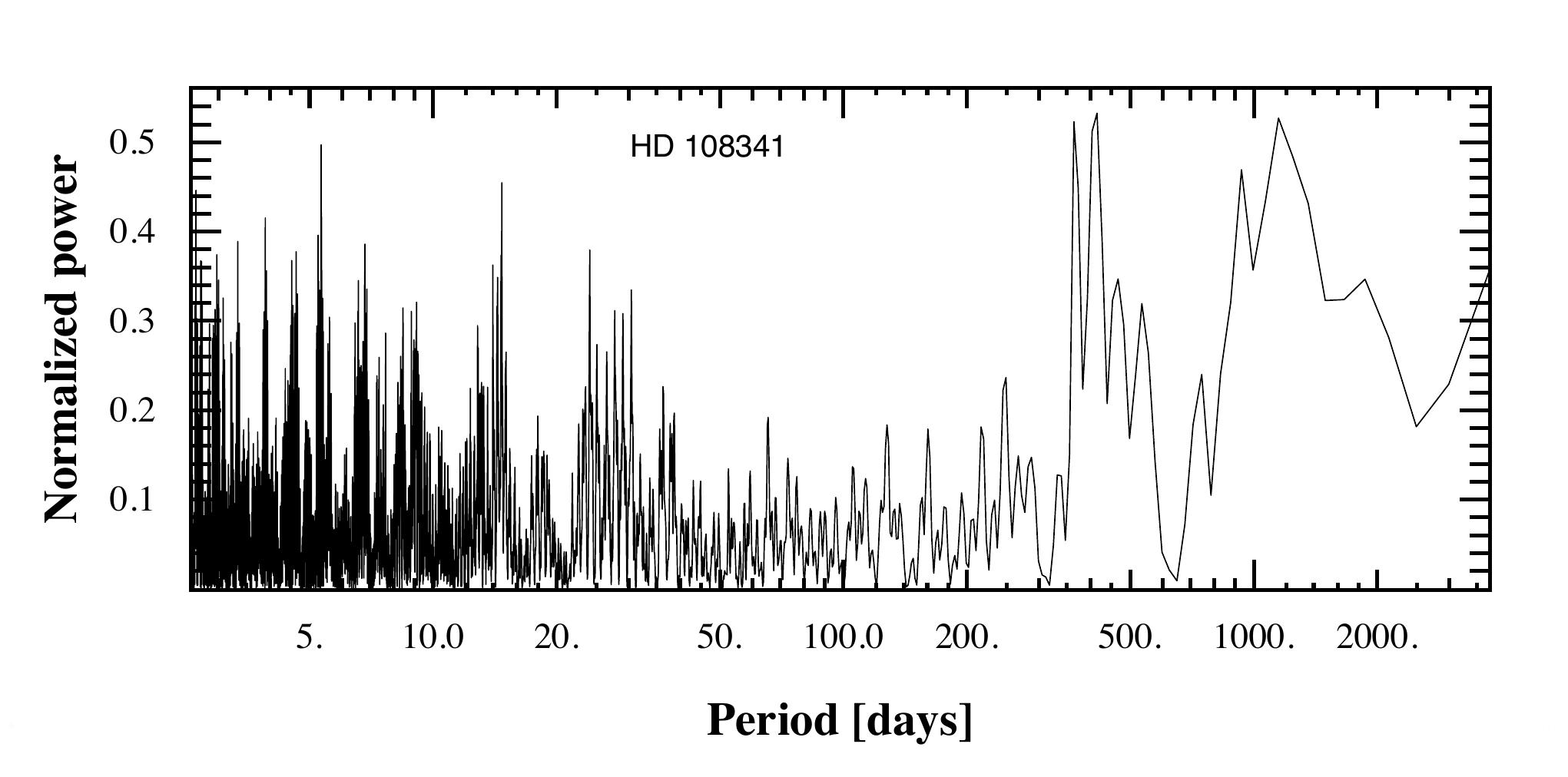,width=0.48\textwidth}
\epsfig{file=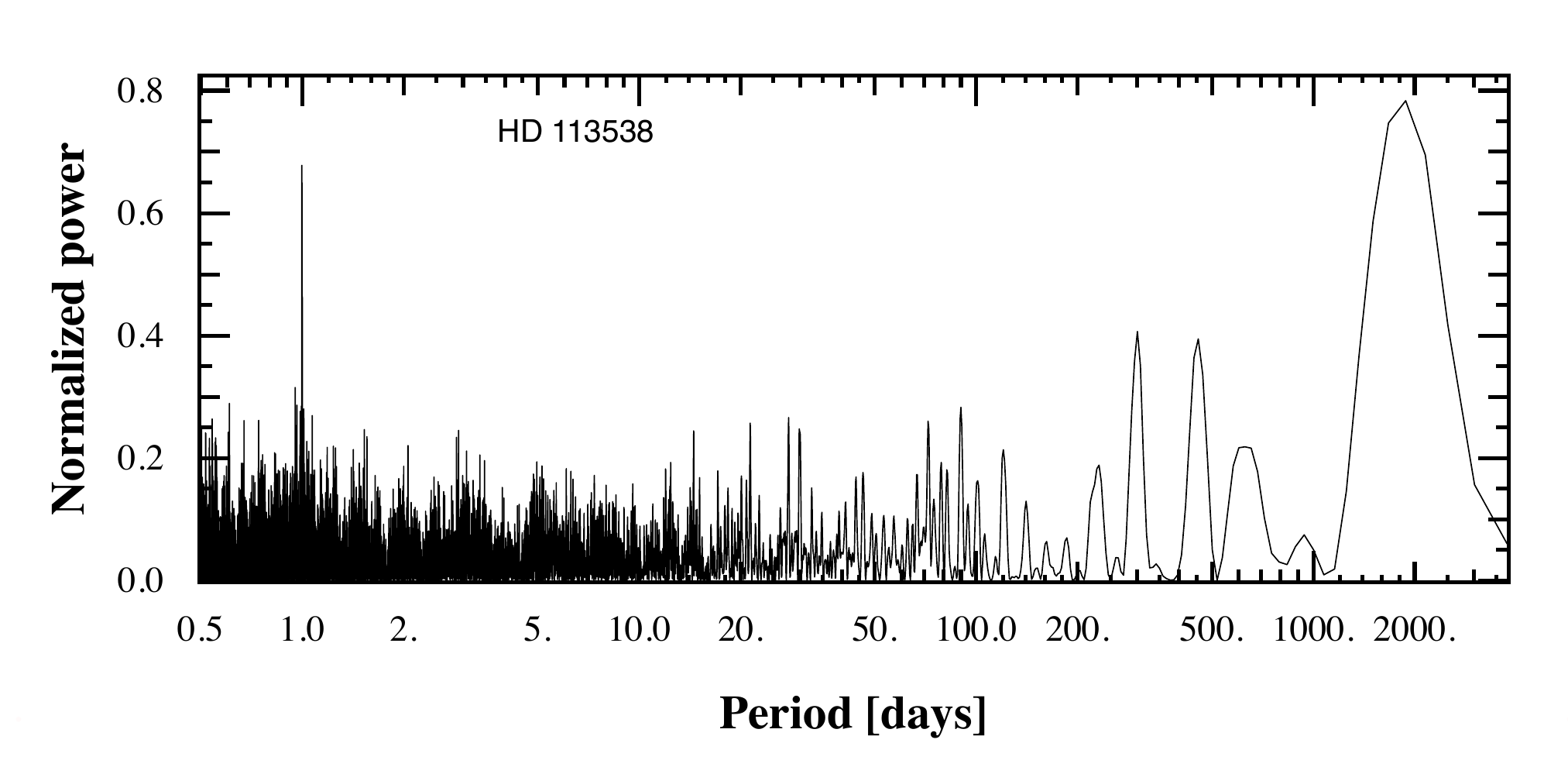,width=0.48\textwidth}\epsfig{file=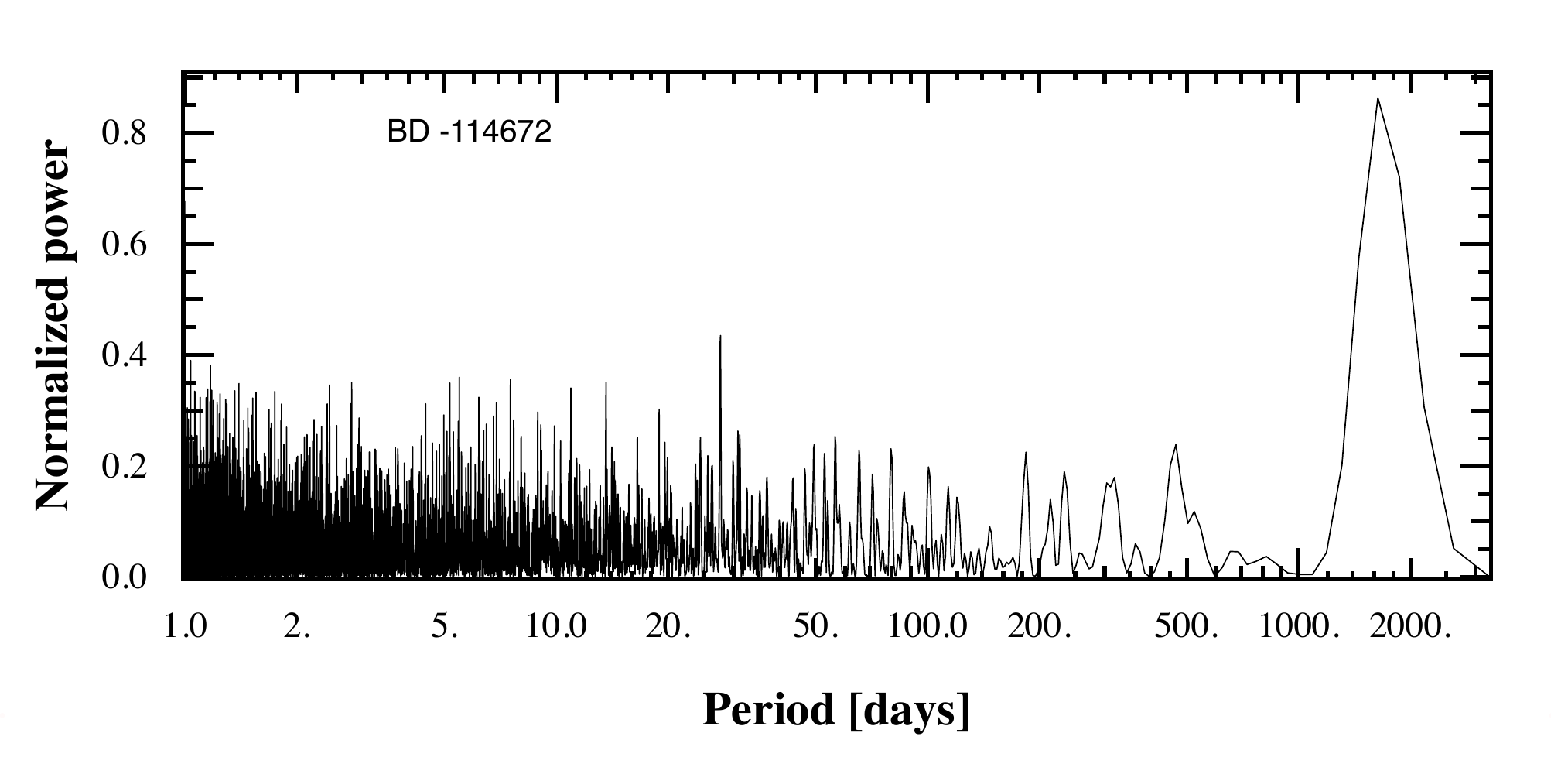,width=0.48\textwidth}
\caption{Periodogram of ten-year HARPS observations.}
\label{gls}
\end{figure*}

\section{Radial velocity data and proposed solutions}
\label{planet}

Radial-velocity data time series were secured with the spectrograph HARPS, which is mounted full time on the 3.60 m telescope in La Silla, ESO, Chile. The spectrograph is kept in the vacuum under strict temperature control and is fed by an optical fiber with a one-arcsec on-sky aperture. HARPS \citep{mayor03} was used in the HAM mode, allowing the highest resolving power of 115,000, without simultaneous lamp calibration. The observations were conducted in the context of the volume-limited program \citep{locurto10}, using a precision of about 2 \ms\ per measurement in the search for systems containing giant planets. Journals of the observations and RV values are reported in the online Tables 4 to 9. We secured 70, 99, 46, 24, 75, and 40 individual measurements for \a, \b, \c, \d, \e, and \f, respectively. Typical S/N values range from 20 to 50, corresponding to individual RV uncertainties of 1 to 3 \ms. 

The data were processed with the HARPS pipeline, including activity-indicator estimates. The radial velocities were obtained by cross-correlation of the spectra with a numerical mask, as first described in \citet{baranne1996}. The mask was adapted to the spectral type of the star, that is,
 a K5 mask for \a, \d, \e, and \f, and a G2 mask for \b and \c. The bisector span, a reliable activity indicator, was measured using the cross-correlation profile (see method in  \citet{queloz01}), and its errors were estimated to be twice the RV errors. The CaII H and K chromospheric emission was measured in all spectra (as described in, e.g., \citet{lovis11}) and calibrated to the Mount Wilson $S$ value and, when appropriate, the $\log R'_{\mathrm{HK}}$.

The RV time series were then analyzed with the {\sc yorbit} software based on a genetic algorithm \citep{segransan11}, and final parameter sets were calculated using a Markov chain Monte Carlo analysis with Metropolis hasting and looped over 500,000 iterations.

The periodograms of the RV observations are shown on Fig. \ref{gls} for all targets. Multiple daily and yearly aliases are visible in addition to the signal peak(s). In the following, we discuss our analyses and interpretation of the RV time series for each individual target.

\subsection{HD 103720}
The 70 HARPS measurements of star \a show a variability of 100 \ms ; individual measurements have an average uncertainty of 2 \ms. Although the star is active, with an average $\log R'_{\mathrm{HK}}$  of -4.48 (or $S$ of 0.55), and shows visible variability in the bisector span and CaII individual measurements (Figs. \ref{bis} and \ref{smw}), it is unlikely that activity alone produces such a high RV variation. From calibrations derived between CaII and RV activity jitter \citep{santos00}, one expects a jitter of 9 \ms\ for a star such as \a, a factor 7 smaller than the RV variability. In addition, neither the bisector span nor the CaII index shows variations at frequency similar to the 4.5-day period of the velocities, and these quantities are not correlated with velocity. The photometric variability of this star is reported in the VSX\footnote{http://www.aavso.org/vsx/} database as characterized by a 17-day period and a 0.05 magnitude amplitude \citep{watson}.

\begin{figure}
\epsfig{file=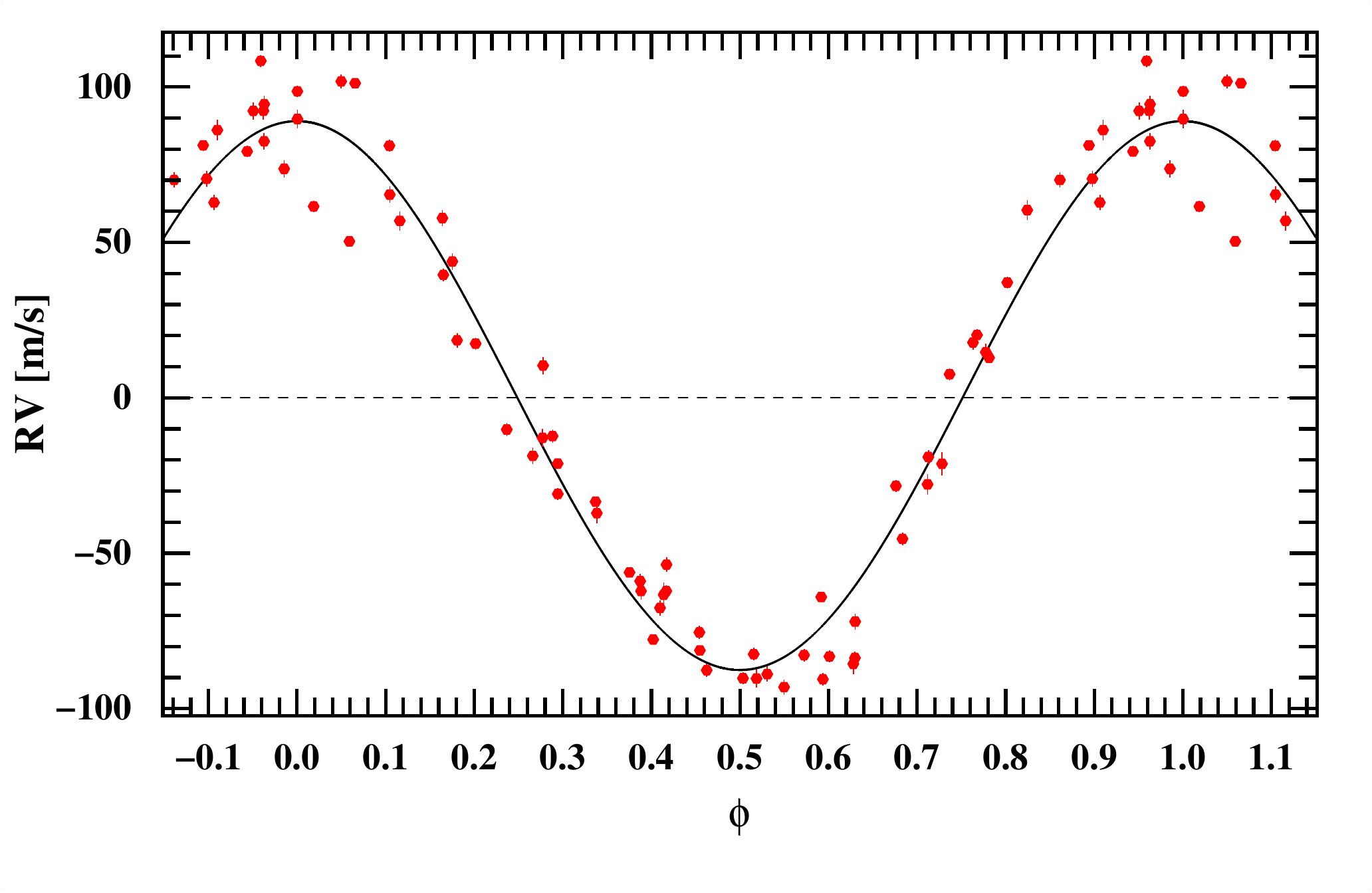,width=8cm}
\epsfig{file=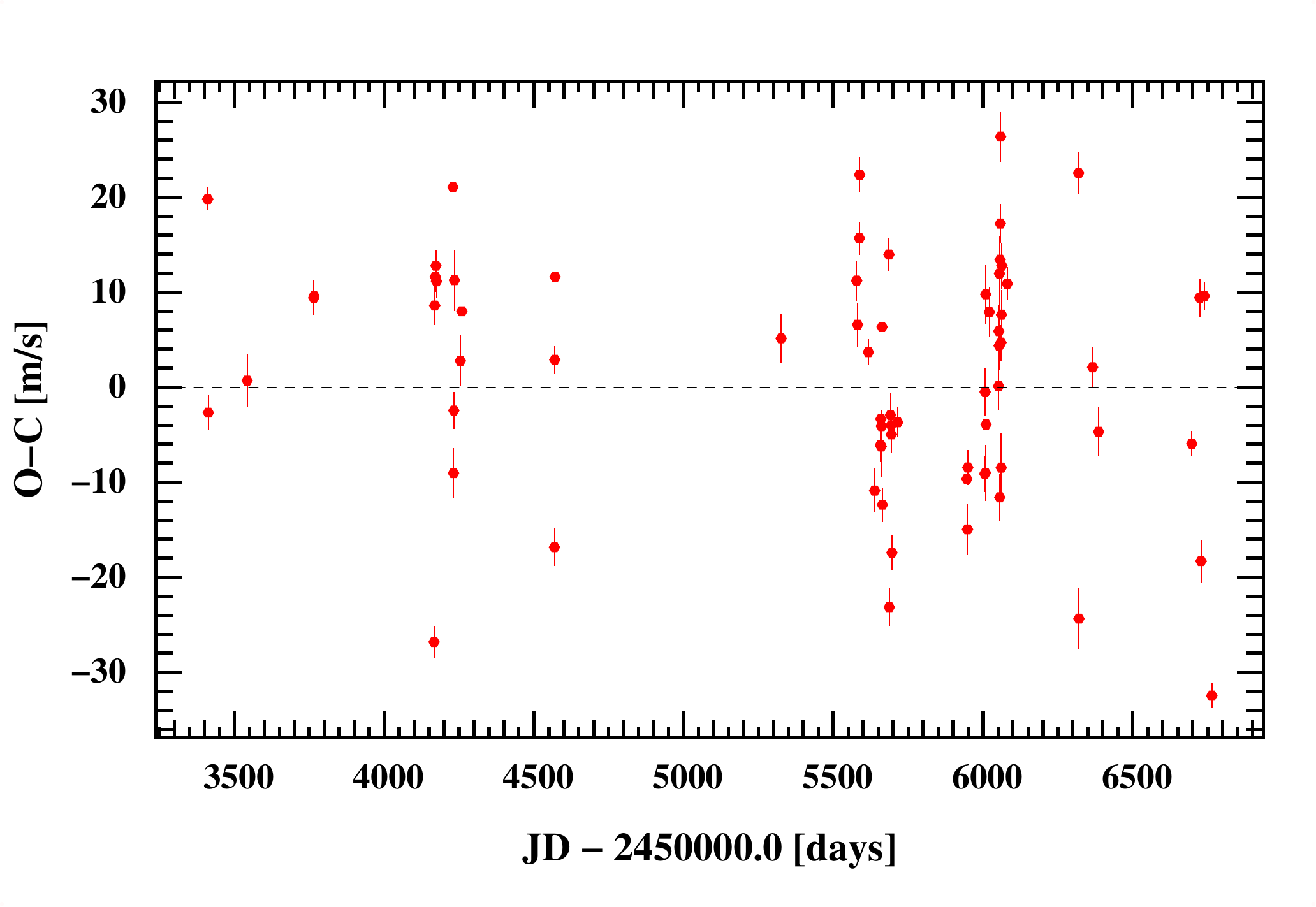,width=8cm}
\caption{Radial-velocity measurements of \a obtained with HARPS against the planet orbital phase together with the best-fit solution  (top) and the residuals over time (bottom). }
\label{rva}
\end{figure}

When a Keplerian is adjusted to the RV data of \a, the RV variations are compatible with the presence of a sinusoidal wobble of amplitude 89 \ms\, that would be produced by a planet of period $P$ = 4.555 days in a circular orbit, and of mass  $M_p$ = 0.62 \mjup. Figure \ref{rva} shows the phase-folded data points together with the best-fit solution and the residuals over time. There is no clear periodicity in the residuals when the main Doppler-induced signal is removed. The residual jitter has an $rms$ amplitude of 11 \ms. It excludes another giant planet of more than 1 \mjup\ in a 70-day circular orbit, and a 4 \mjup\ planet in a 3000 day orbit. Since the residuals are compatible with the expected jitter, they are attributed to spot and plage activity at the surface of the star, while the main signal is safely attributed to a giant planet in orbit. 

The final parameters of the planet are given in Table \ref{sol1}.

With an orbital period of 4.555 days and a stellar radius of 0.73 R$_{\odot}$, this giant planet has a 7\% probability of transiting the disk of its parent star. All system parameters are given in Table \ref{sol1}. The ephemeris of a transit configuration are $T_0$ = 2455388.709$\pm$0.047 and $P$ =4.5557$\pm$0.0001 days. No attempt to search for the transit has been made.

\subsection{HD 564}

\begin{figure}
\epsfig{file=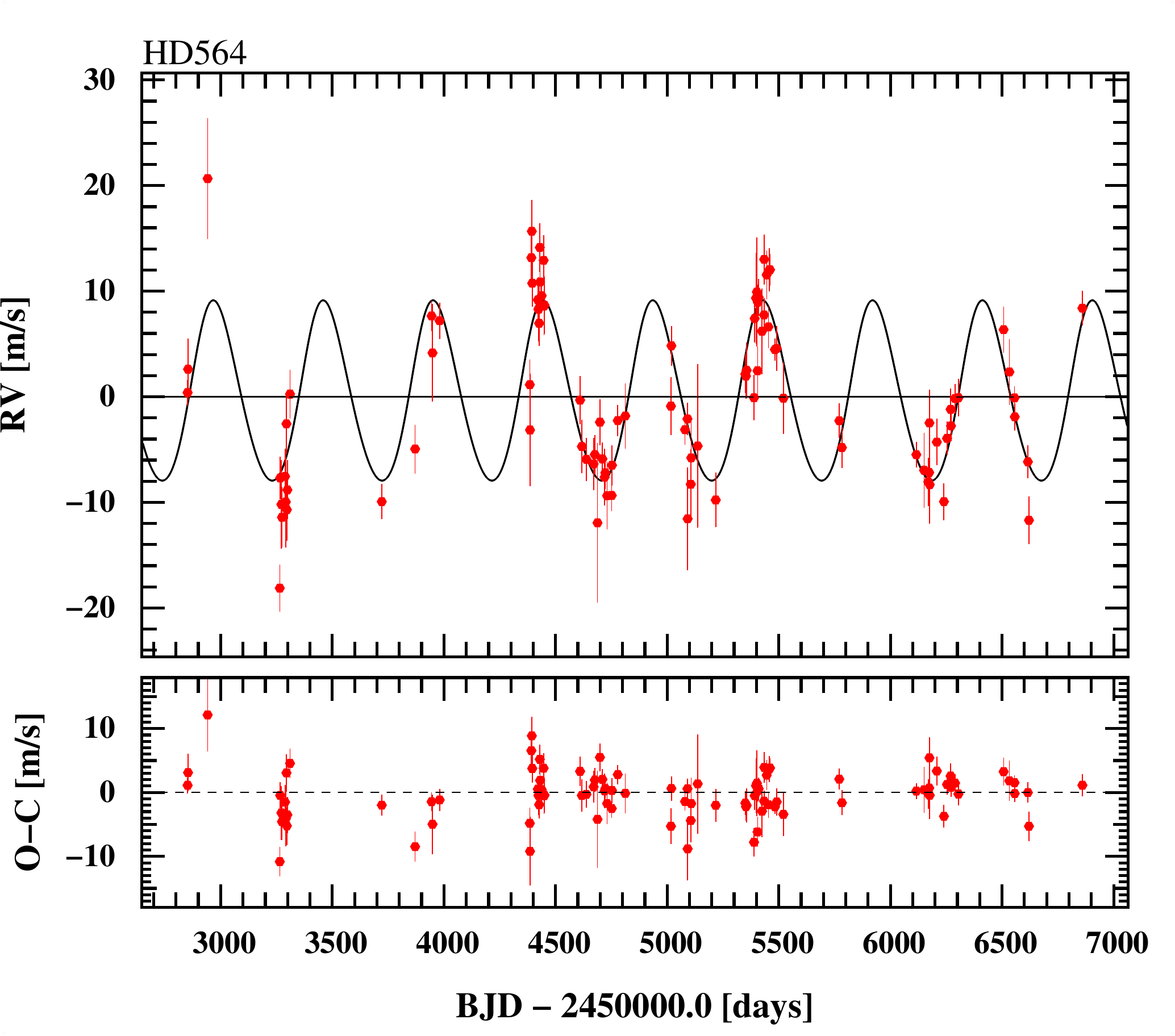,width=8cm}
\epsfig{file=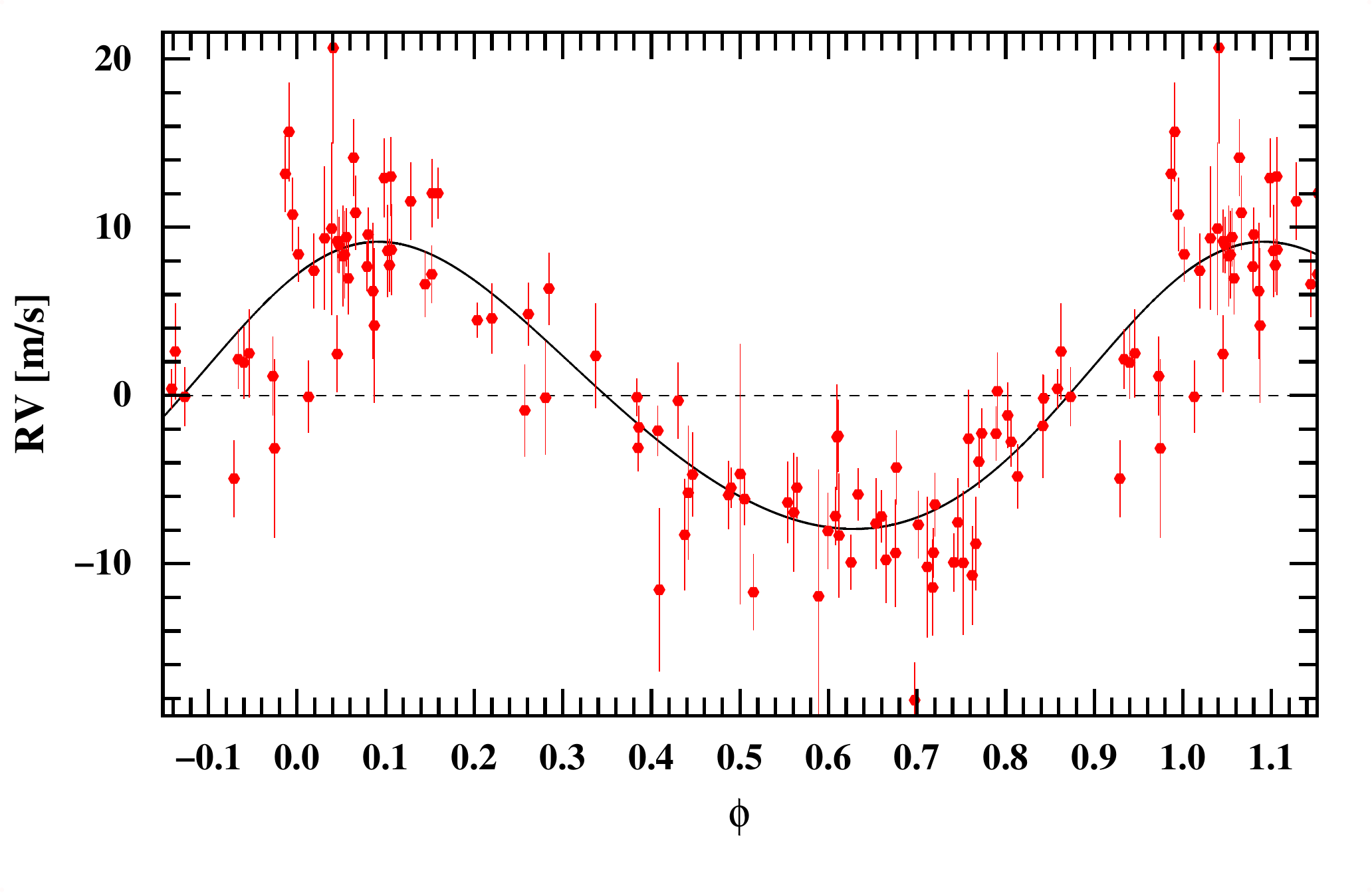,width=8cm}
\centering\epsfig{file=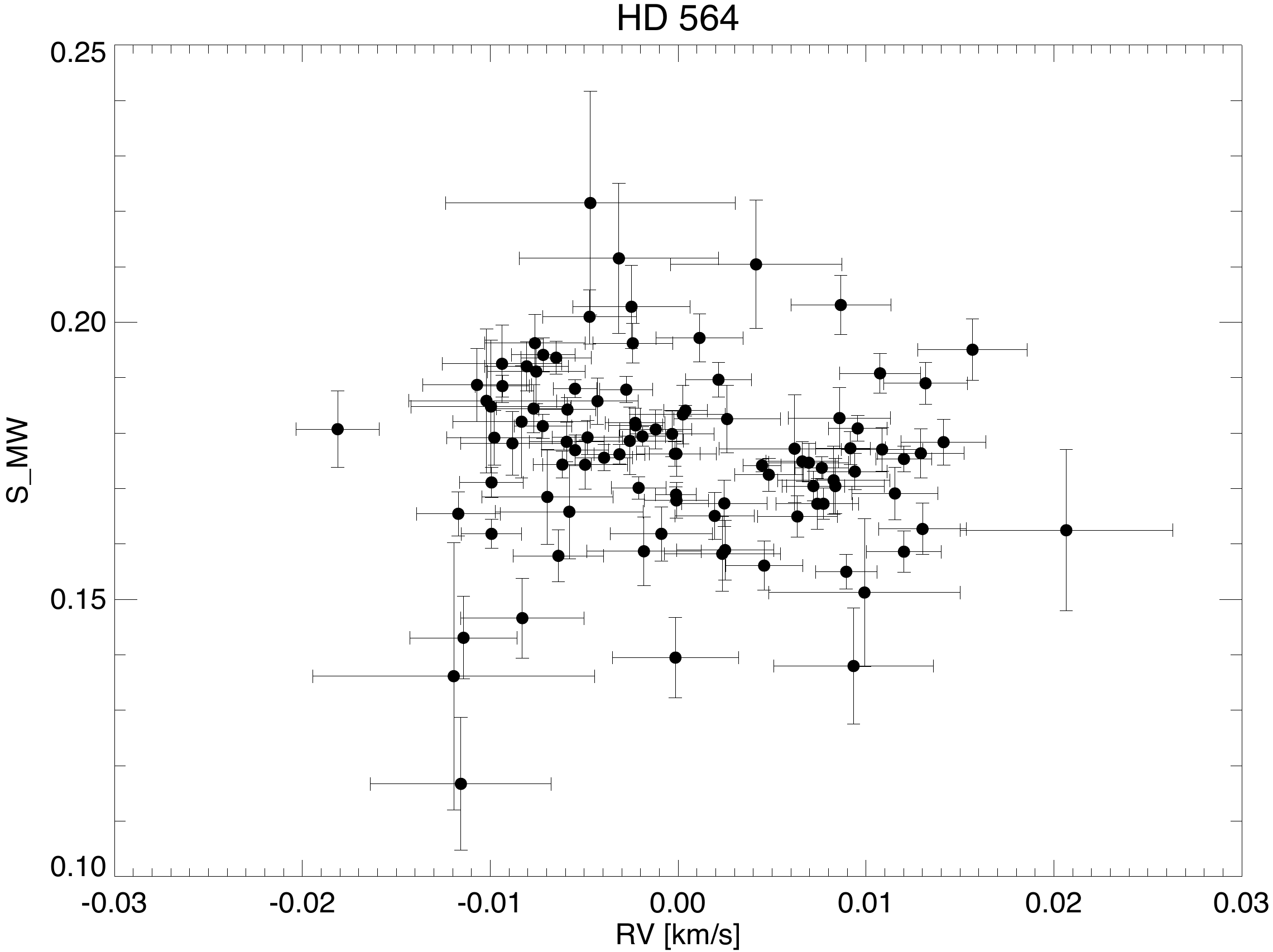,width=7.5cm}
\caption{Radial-velocity measurements of \b obtained with HARPS against time (top) and against the planet orbital phase (middle) together with the best-fit solution. The CaII index is not correlated with the RV variations (bottom). }
\label{rvb}
\end{figure}

The metal-poor star \b has been observed during more than ten years with HARPS, during which 99 data points were collected with an average accuracy of 2.5\ms. The RV measurements show a variability of 20 \ms\ amplitude, or twice the jitter level expected from the chromospheric activity index. 
The CaII index is not correlated to the RV variations (bottom of Fig. \ref{rvb}). The bisector span varies over a similar range as the velocities, but with a Pearson correlation coefficient of 0.04. Even if the star may show some activity despite its low  $\log R'_{\mathrm{HK}}$ value, we are confident that activity alone cannot explain  the RV variations, because i) we cover nine periods and the power in the RV-periodogram peak at 490 days is very strong over the ten years of monitoring, showing a stable signal, ii) activity indices have power peaks spread over a wider range of periods, iii) there is no correlation between velocities and activity indices. 

When a Keplerian is adjusted to the 99 HARPS data points, the best fit has an orbit of 492.3$\pm$2.3 days and a semi-amplitude of 8.79$\pm$0.45 \ms\ (Fig. \ref{rvb}). The $rms$ of the residuals is 2.9\ms\ and the periodogram of the residuals does not show any periodicity. This signal would correspond to a 0.33\mjup\ planet at 1.2 AU from its host star (Table \ref{sol1}). 

\subsection{HD 30669}

\begin{figure}
\epsfig{file=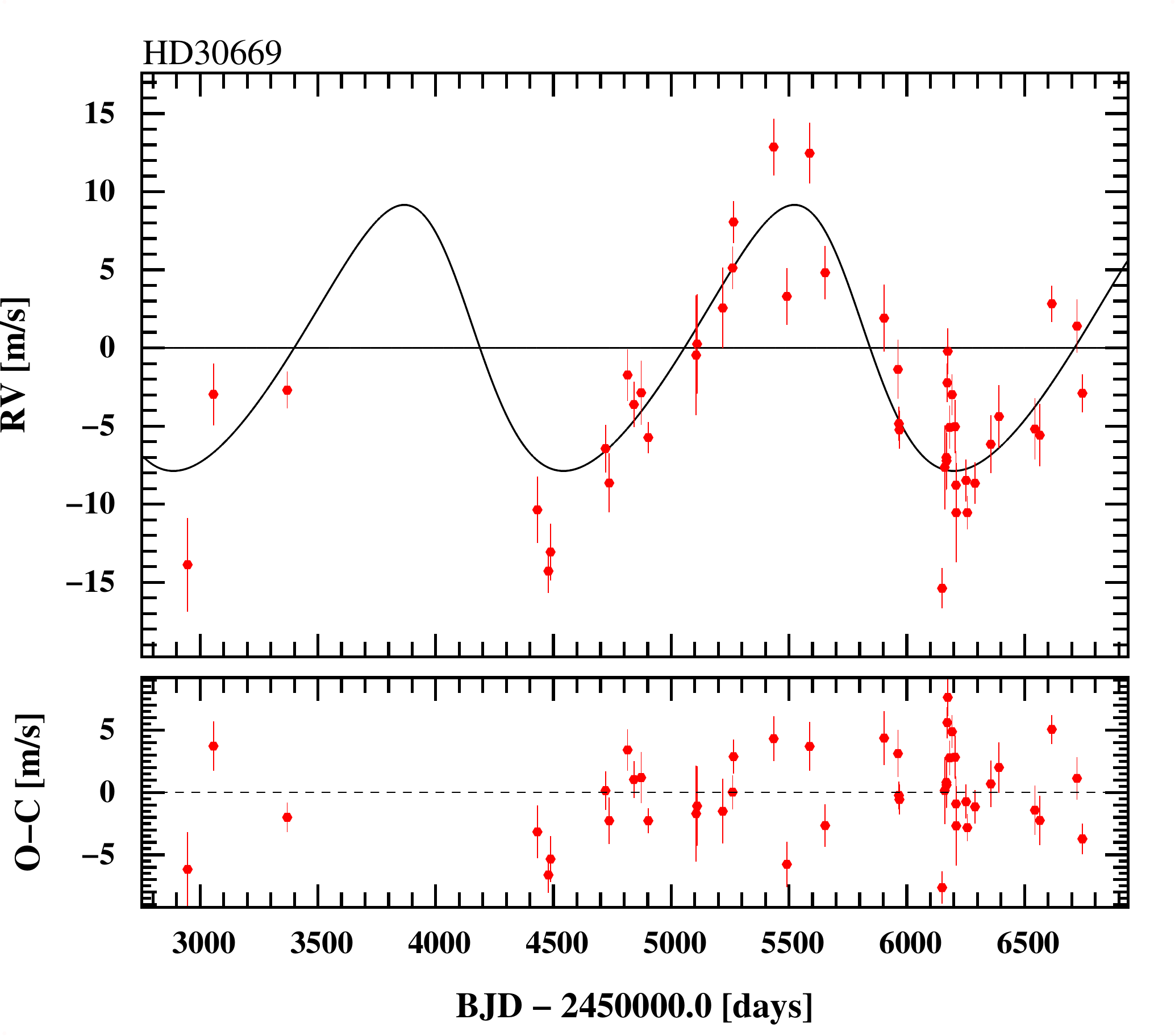,width=8cm}
\epsfig{file=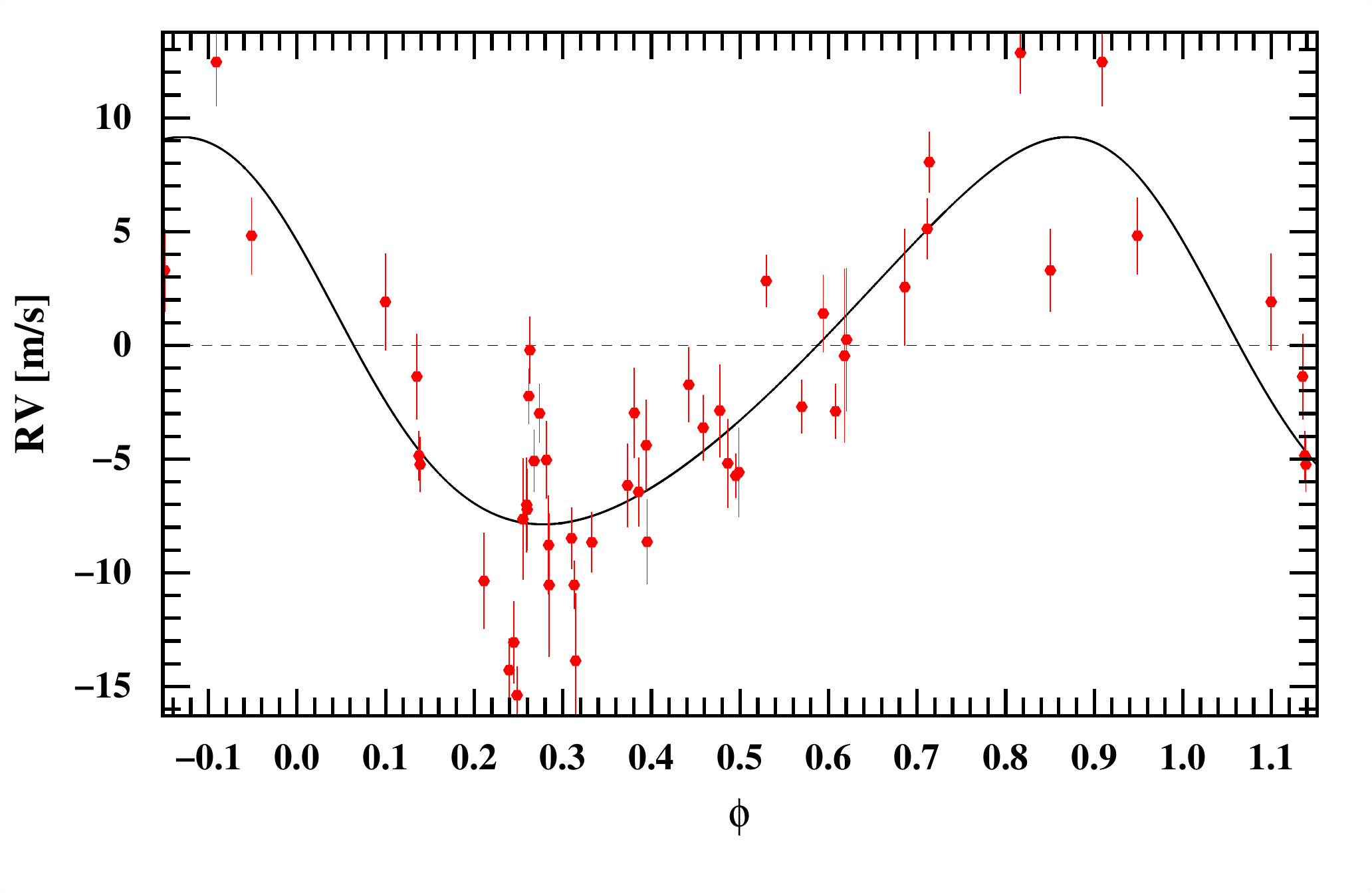,width=8cm}
\caption{Radial-velocity measurements of \c obtained with HARPS against time (top) and against the planet orbital phase (bottom) together with the best-fit solution. }
\label{rvc}
\end{figure}

The star \c\, has been observed 46 times with HARPS over ten years, with an average accuracy of 1.8 \ms. It is the least active star of the sample described in this work, with an expected velocity jitter of about 7 \ms\ due to stellar activity \citep{santos00}. 
The strongest peak in the periodogram of the RV time series lies at about 1680 days. This peak is not present in the periodograms of the CaII index or of the bisector span. Since activity is not an issue, a Keplerian was adjusted to the data, and the best-fit solution corresponds to a signal at a period of 1684$\pm$61 days and semi-amplitude 8.6$\pm$1.1\ms\ (Fig. \ref{rvc} and best-fit model parameters in Table \ref{sol1}). The $rms$ of the residuals is 3.6\ms. The periodogram of the residuals shows a weak peak at 150 days that could be due to another companion. The $rms$ of the residuals drops to 1.7\ms\ when this second signal is adjusted. Since activity is still expected to have a few \ms\ amplitude jitter, we did not consider this additional companion as significant based on the present data set. It could rather be an activity artifact, with a semi-periodic signal at $\sim$ times the rotational period.

\subsection{HD 108341}

\begin{figure}[h]
\epsfig{file=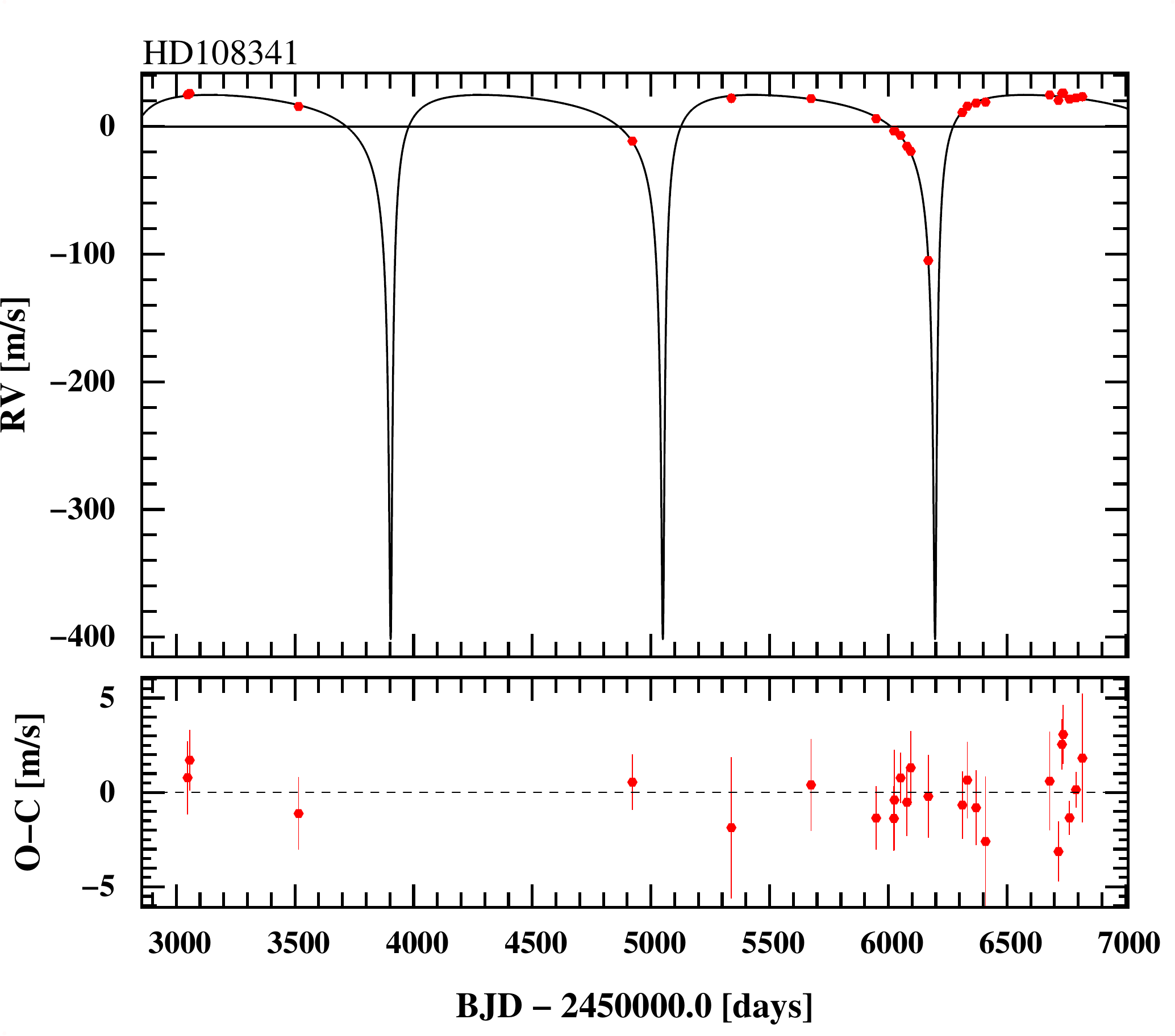,width=8cm}
\caption{Radial-velocity measurements of \d obtained with HARPS against time (top) and the residual to the best-fit model (bottom). }
\label{rvd}
\end{figure}

\begin{figure}[h]
\epsfig{file=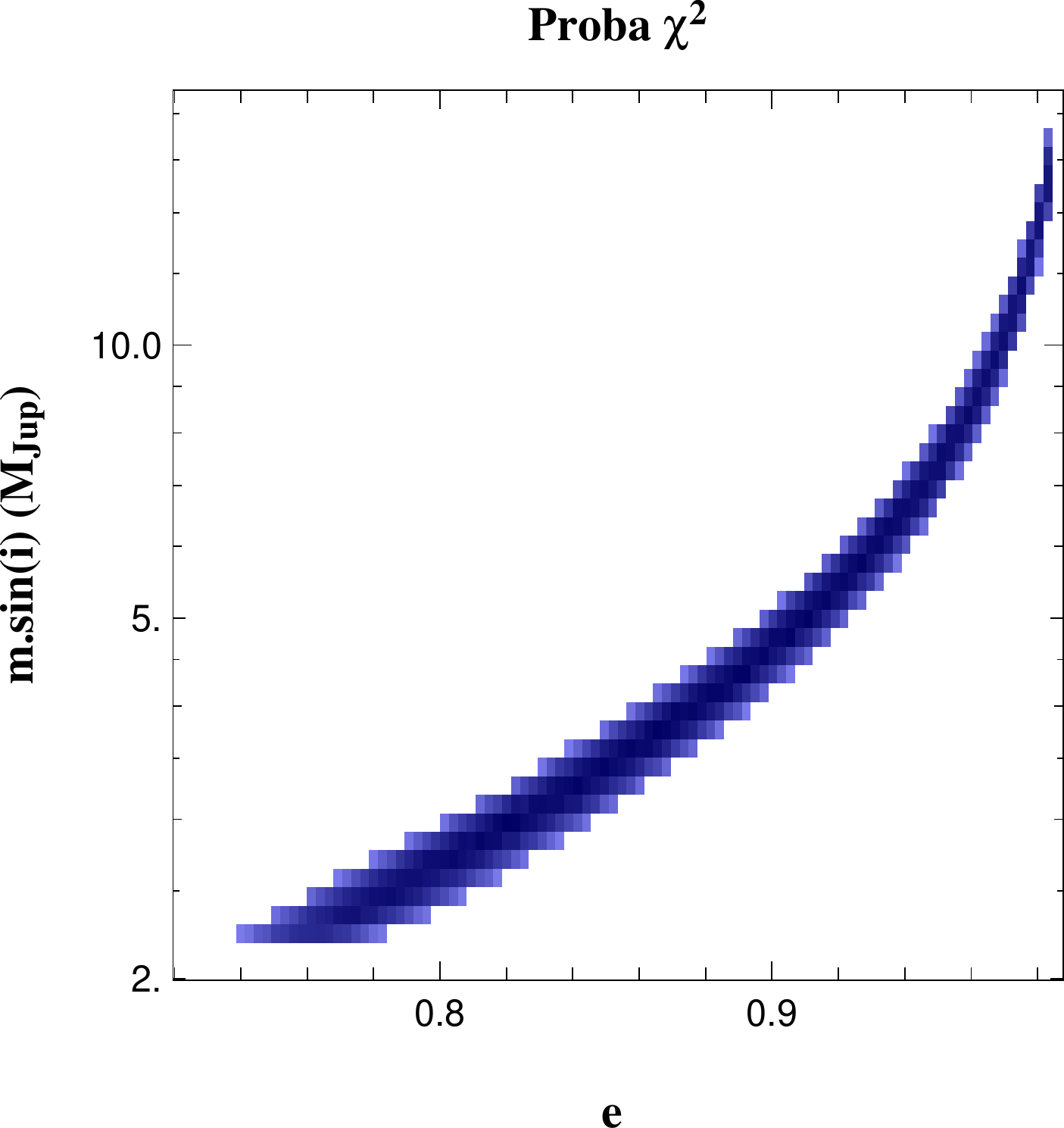,width=8cm}
\caption{$\chi^2$ map of the companion mass and eccentricity obtained from the RV data set of \d. }
\label{chi}
\end{figure}

The K star \d was observed 24 times with HARPS over more than ten years, with individual RV errors of 1.9\ms\ average. The star has a low activity index, and neither the bisector nor the CaII index varies with velocity. The RV variations of \d span from -110\ms\ to +20\ms, featuring the signature of a possible eccentric companion. When adjusted by a Keplerian orbit, the best-fit solution has a semi-amplitude of 144$^{+311}_{-66}$ \ms\  and a period of 1129 days (Fig. \ref{rvd}).
When this signal is removed, the $rms$ of the residuals is 1.5\ms, which is at the photon noise limit. Although the periastron passage is still undersampled, resulting in a large, asymmetric uncertainty in the final parameters (Table \ref{sol1}), we are confident that the detection of a giant planet in an eccentric orbit at about 2 AU semi-major axis is definite. The large uncertainty on the semi-amplitude is due to this incomplete periastron coverage, resulting in a two-peaked posterior distribution of most orbital parameters. The correlation between eccentricity and companion mass, as allowed by the data, is shown in Fig. \ref{chi}. At this stage, the companion mass is only poorly constrained (2.3 to 6.9 \mjup\ in the 68.3\% confidence interval).
Additional HARPS measurements, especially in August 2015 during next expected periastron passage, will decrease the error bars on the mass, period, and eccentricity.

\subsection{HD 113538}

We present here an update of the analysis of HARPS measurements of \e, in which two planets were already reported in \citet{moutou11}. We have now collected 75 HARPS measurements of \e, spanning 3700 days, while there were only 29 (over 2268 days) in the previous study and much uncertainty remained on the parameters of these companions. The star is metal-poor and of late spectral type; the latter explains why $\log R'_{\mathrm{HK}}$  could not be estimated (nor the expected activity jitter), the stellar parameters were beyond reliable calibration boundaries. However, the behavior of the CaII $S$ index as a function of time shows signs of activity and possible long-term cycling (Fig. \ref{smw}). The chromospheric activity of the star currently increases after a minimum about four years earlier. 

\begin{figure}[h]
\epsfig{file=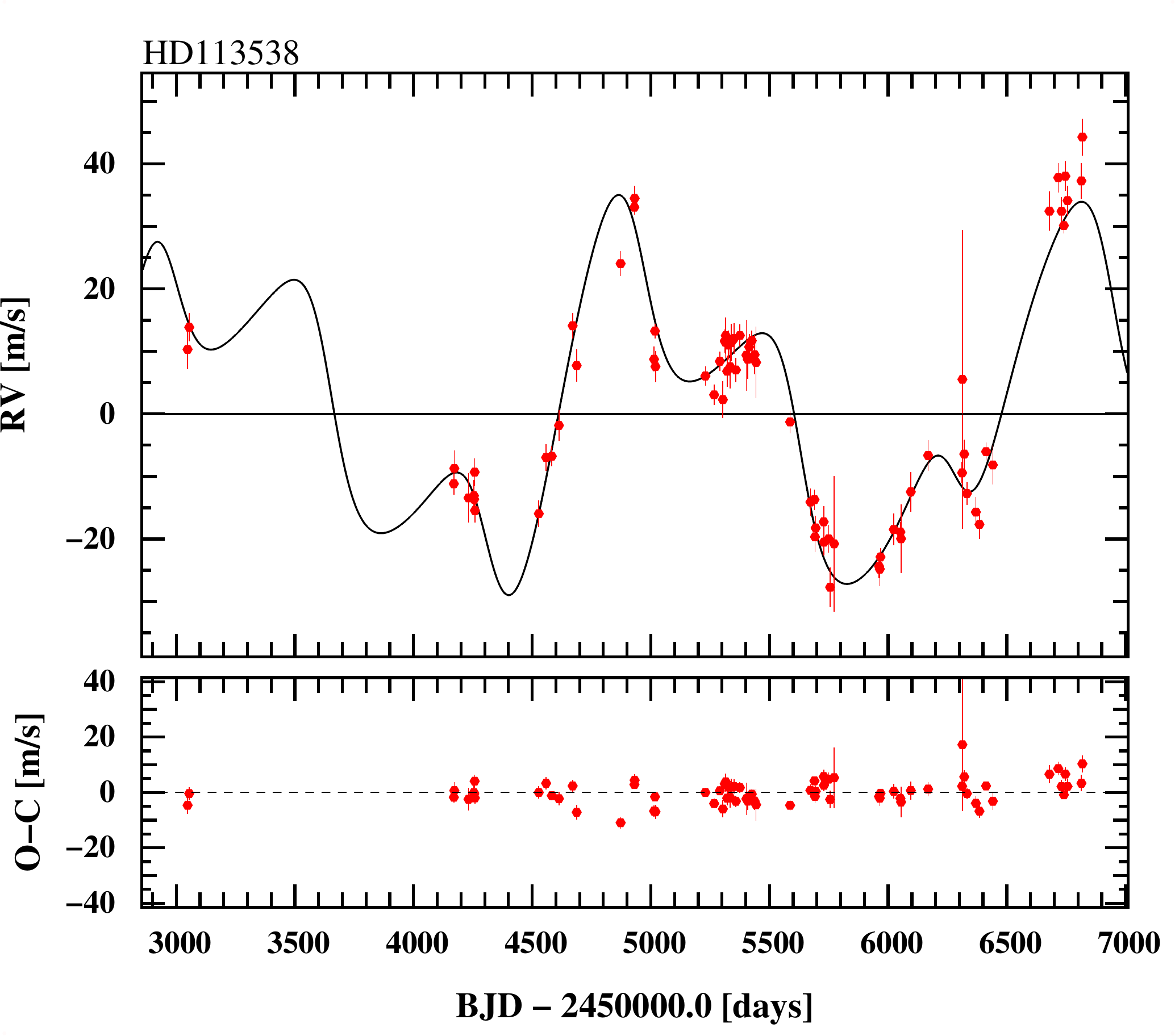,width=8cm}
\epsfig{file=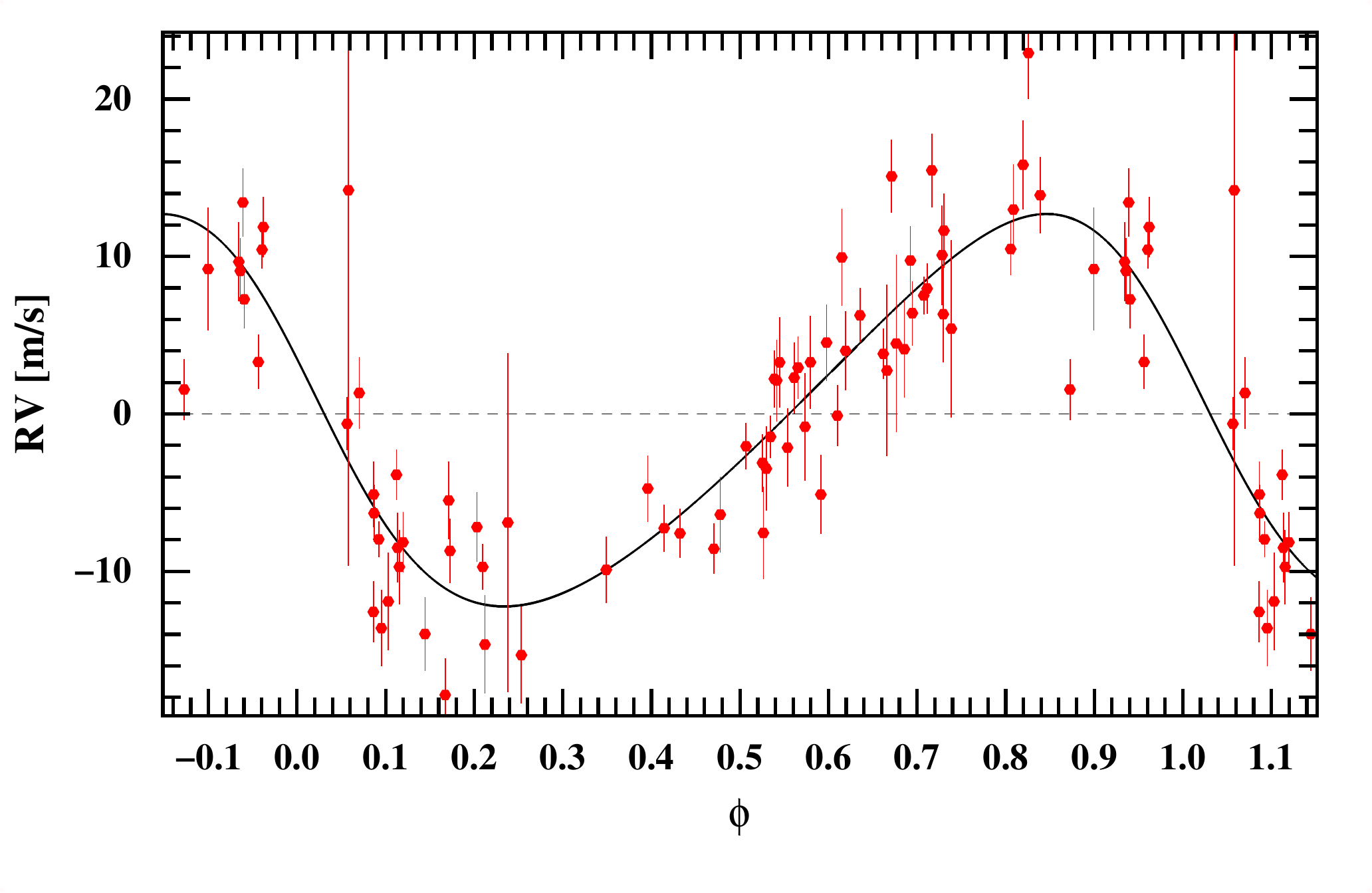,width=8cm}
\epsfig{file=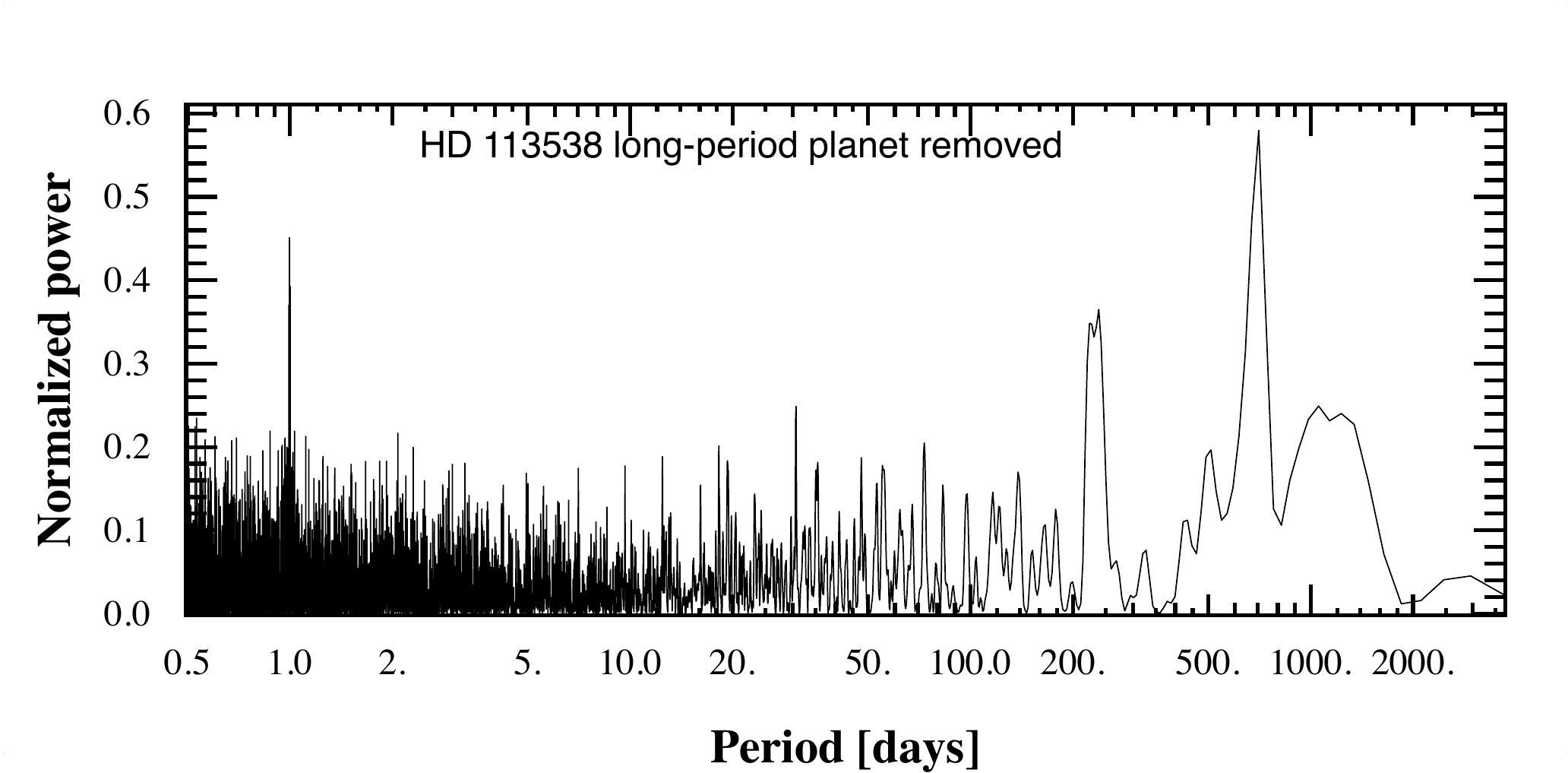,width=8cm}
\epsfig{file=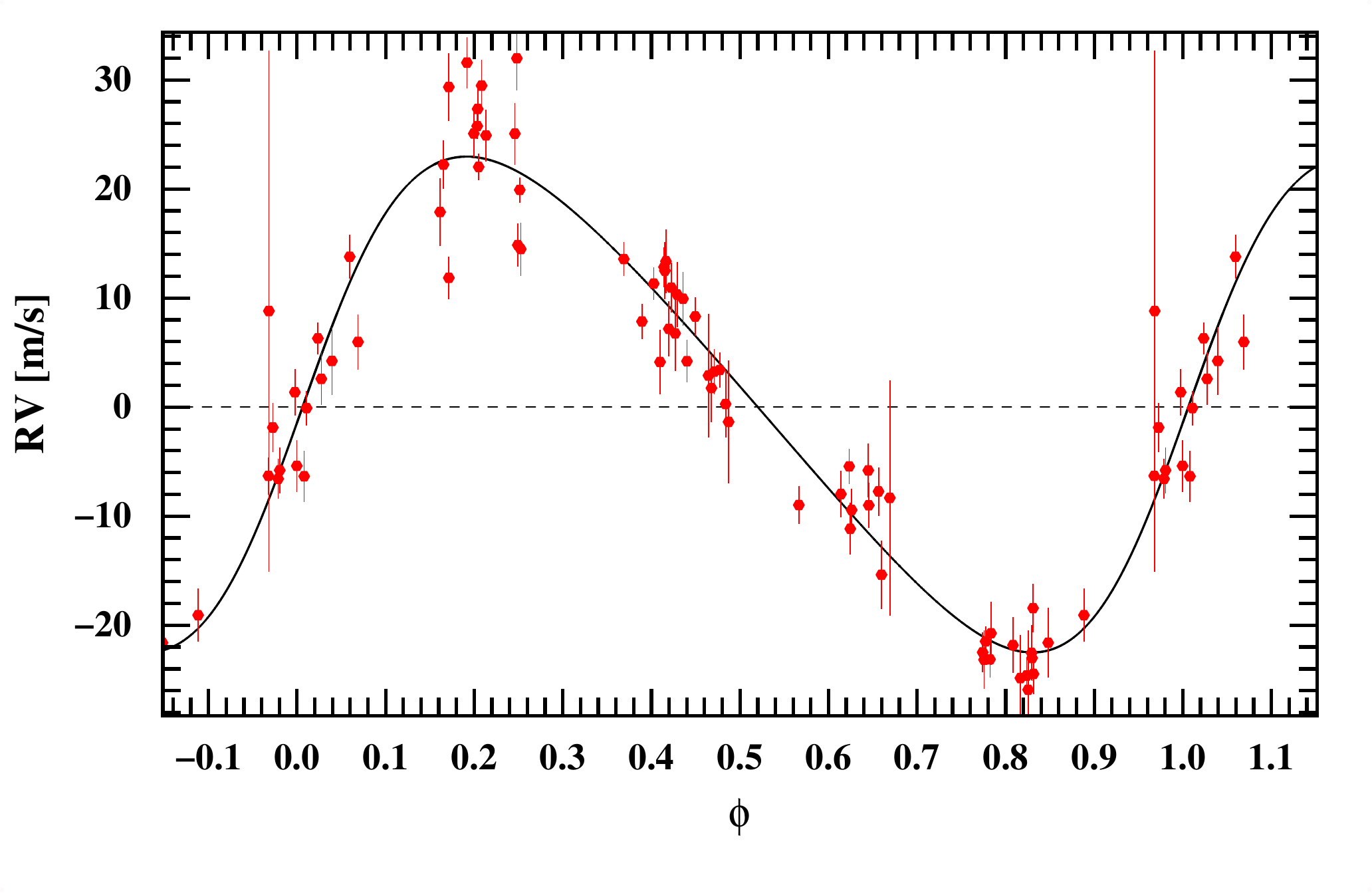,width=8cm}
\epsfig{file=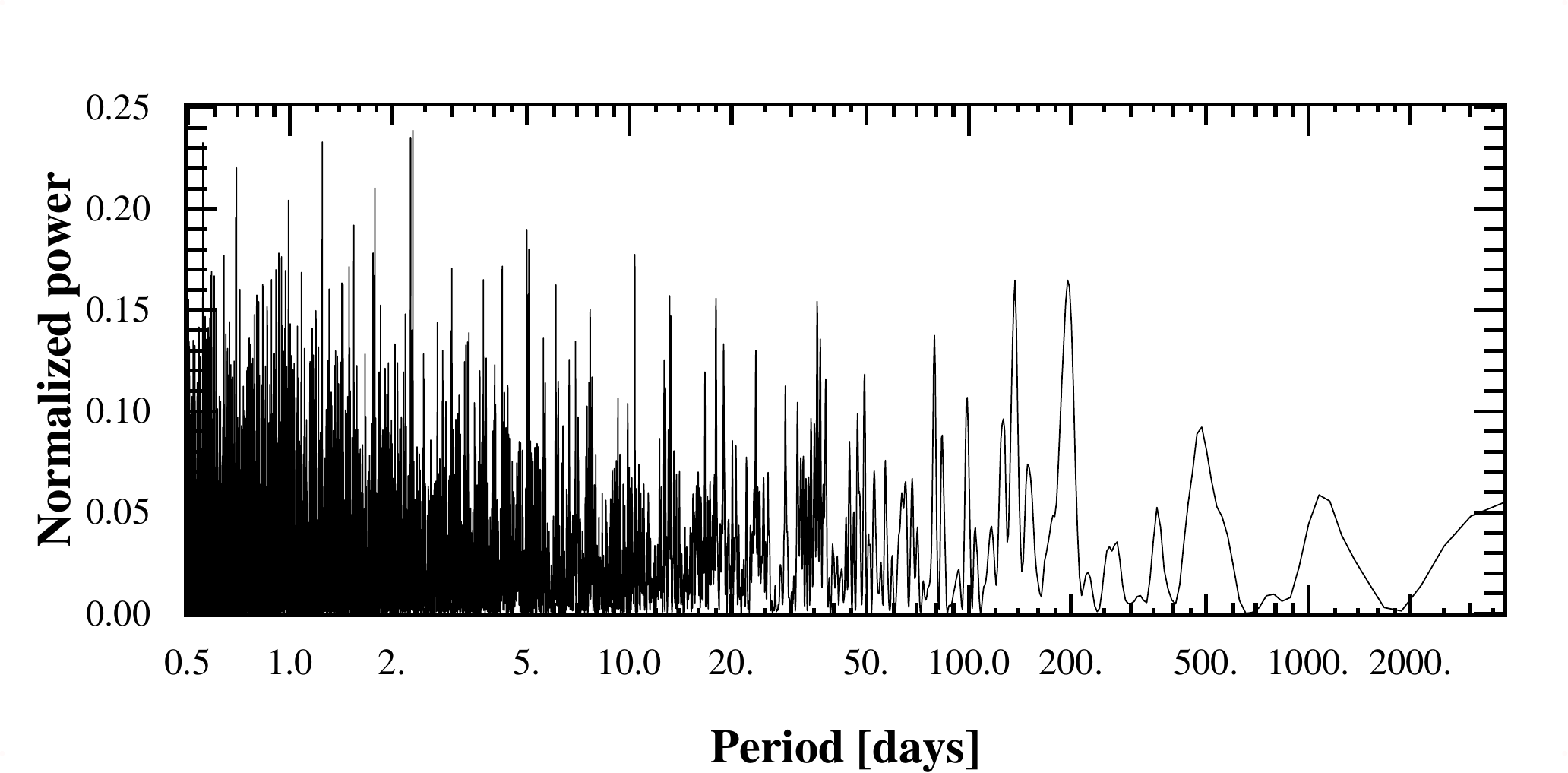,width=8cm}
\caption{Radial-velocity measurements of \e obtained with HARPS against time (top) and versus phase  for both planets (second line: 660-day period planet; fourth line: 1800-day period planet). The periodograms of the residuals after 1 (2) planet(s) are removed are shown in the third and fifth lines.}
\label{rve}
\end{figure}

As in 2011, the current  RV data set of \e is again better adjusted by the combination of two Keplerian curves. When adjusted with a first Keplerian, the best-fit corresponds to a period of 1870 days and a semi-amplitude of 24\ms; the $rms$ of the residuals is 7\ms. The periodogram of the residuals shows a strong and well-defined peak at 660 days and another weaker peak at about 250 days. This latter peak - which lies close to the yearly alias of the former - disappears from the residuals when a two-Keplerian fit (including the 660 day signal) is applied. The $rms$ of the residuals is 3.4\ms after the two planet signatures are removed. The periodograms when one and two planet signals are removed are shown in Fig. \ref{rve} together with phase-folded RV series and RV against time plots. The final parameters of the planets are given in Table \ref{sol2}. Compared with the earlier solution proposed by \citet{moutou11}, they are also affected by the update on the stellar mass (see Sect. 2).

Almost four additional years of continuous monitoring of this star have finally deeply changed the system solution since the first attempt by \citet{moutou11}. The new parameters of the system are better constrained. However, with an increase in the chromospheric activity of the star, it is not clear whether the coming years will have optimal conditions to pursue observations on this target. The increasing level of activity may impact the tenuous signal detected from these planets and hamper an even better constraining of their parameters. We made the experiment of removing the last 20 data points, which correspond to the activity increase. It has the effect of reducing the longest period (to 1584$\pm$30 days) and slightly changes the position of the second, lower-amplitude and shorter-period signal near 600 days. The final solution achieved on this data set therefore still depends on the specific data set used. With a span of the measurements so close to twice the period of the longest planet (3770-day span and 1820-day period) and a second maximum observed at JD$\sim$2456700, we are confident that our determination using the full data set is currently the most accurate. But another update in a few years would be profitable since there is room for improvement in the determination of the planetary and orbital  parameters. However, stellar activity may be an issue as its level is currently increasing.

\subsection{BD -114672}

We have doubled the number of HARPS measurements of star \f since the first observation reported by \citet{moutou11} and updated the stellar mass (Sect. 2). The new data set spans 3271 days and is composed of 40 measurements, with an average uncertainty of 1.8\ms. In the previous analysis, we had stayed conservative as the combined analysis of the activity and RV data was inconclusive. There was a non-negligible correlation between the cross-correlation function parameters and the CaII index, and the bisector was anticorrelated with the velocities. Since the time span of the observations was then also too short compared with the possible period and because activity potentially was a problem, we proposed
conclusions of both a long-distant planet and a magnetic cycle. 

\begin{figure*}[h]
\vspace*{-1.2cm}
\epsfig{file=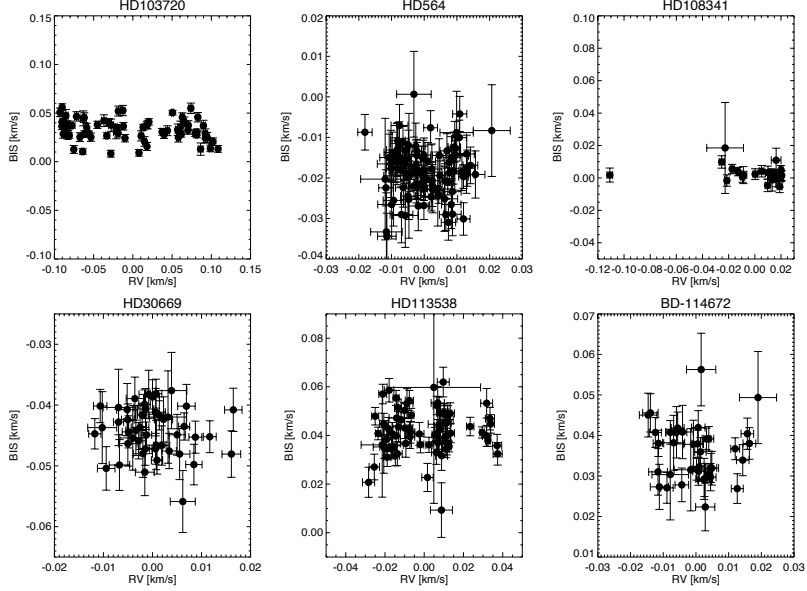,width=0.9\textwidth}
\vspace*{-1.2cm}
\caption{Bisector versus radial velocity }
\label{bis}
\end{figure*}

\begin{figure*}[h]
\vspace*{-1.2cm}
\epsfig{file=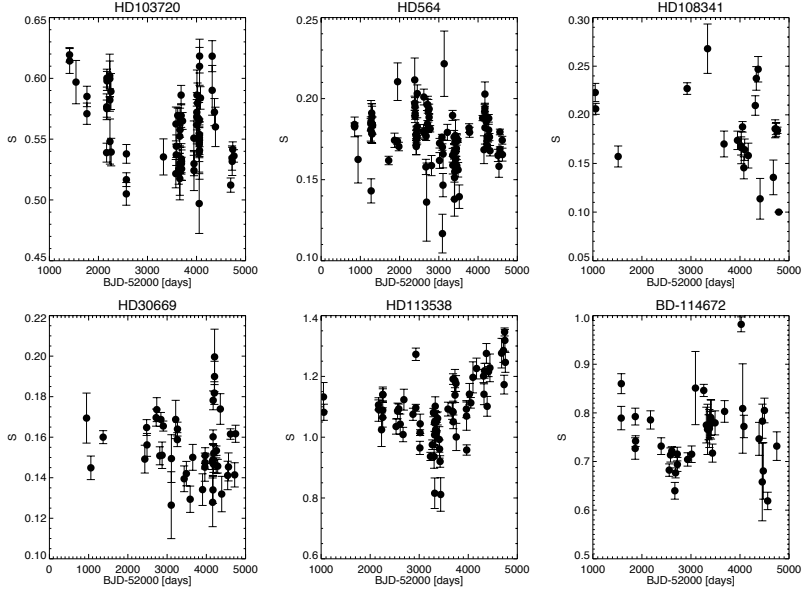,width=0.9\textwidth}
\vspace*{-1.2cm}
\caption{Mount Wilson $S$ activity index versus time}
\label{smw}
\end{figure*}

With the increased data set, the impact of activity has dropped. As shown in Fig. \ref{bis}, the anticorrelation between the bisector span and the radial velocity is not visible anymore. The evolution of the CaII $S$ index with time (Fig. \ref{smw}) is now constant. Finally, the correlation between this index and the FWHM of the cross-correlation function (as evoked in \citet{moutou11}) is weaker, with a coefficient decreasing from 72\% to 50\%. 

\begin{figure}[h]
\epsfig{file=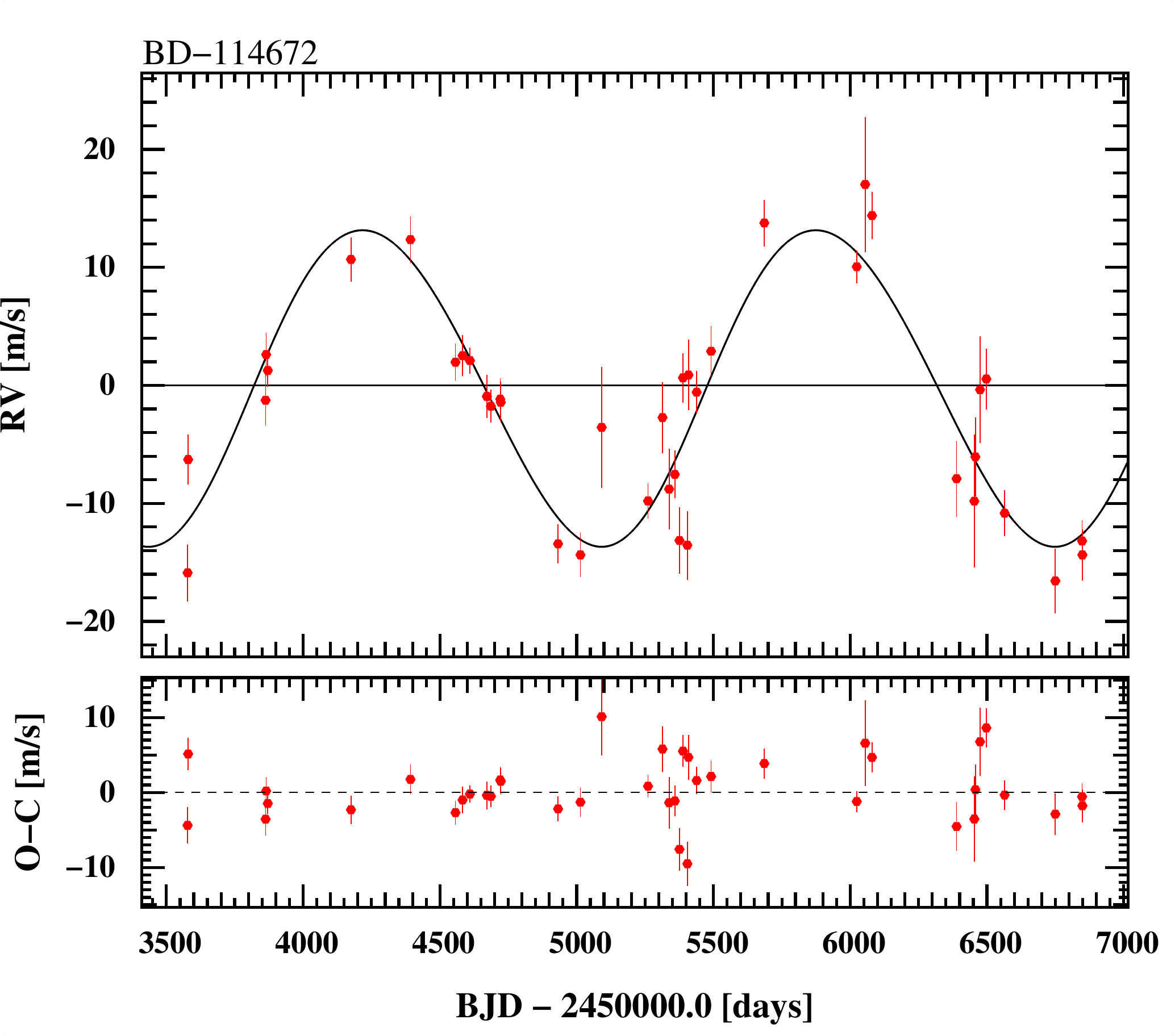,width=8cm}
\epsfig{file=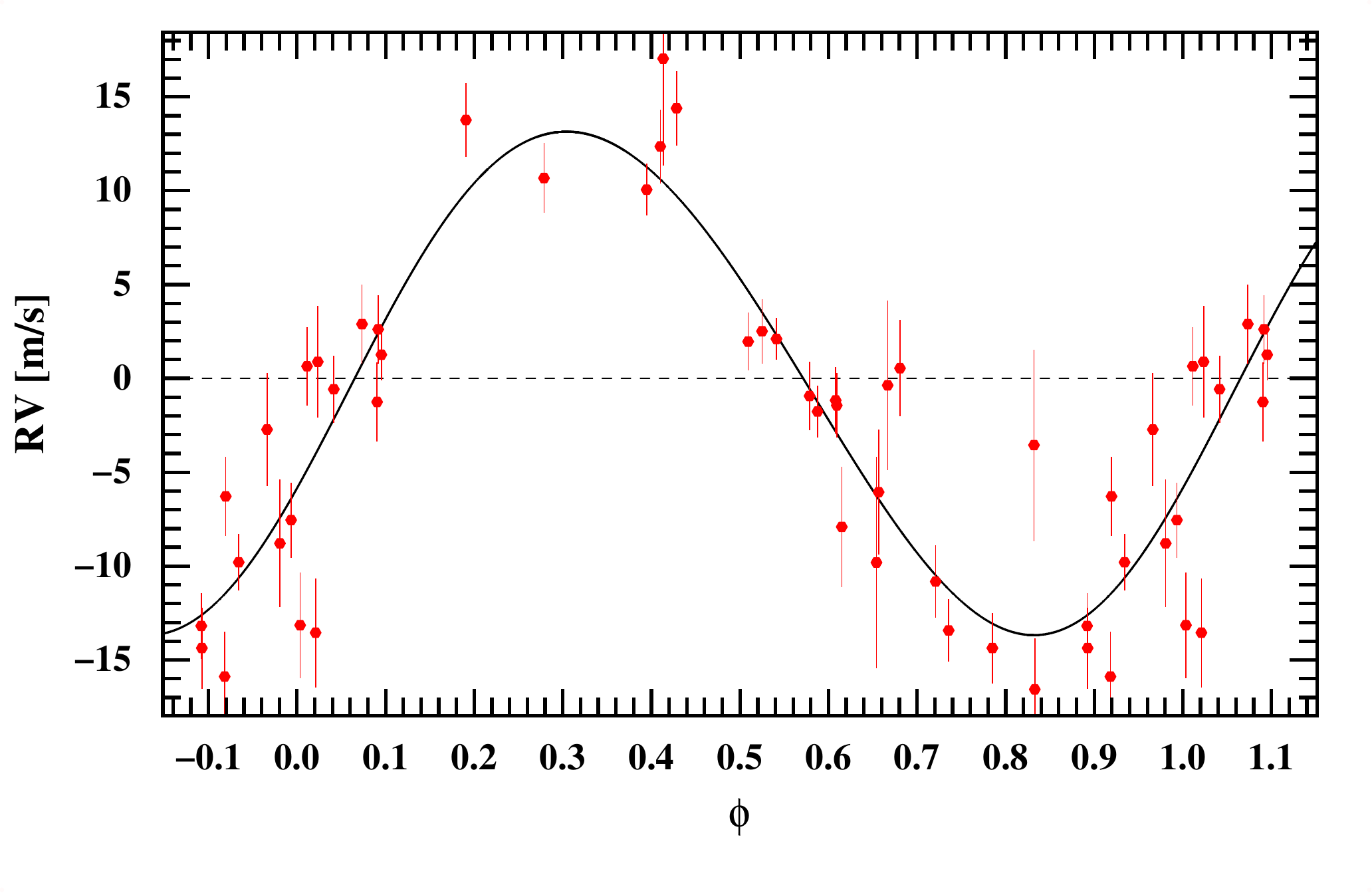,width=8cm}
\caption{Radial-velocity measurements of \f obtained with HARPS against time (left) and phase (right). }
\label{rvf}
\end{figure}

The planet scenario, in turn, has gained significance. The peak in the periodogram is  more pronounced, and with two cycles just covered, the RV signal is better determined. A Keplerian fit gives a best solution characterized by a period of 1667 days, a circular orbit, and semi-amplitude 13\ms, as shown in Fig. \ref{rvf}. The exact parameters and their errors after the MCMC analysis are given in Table \ref{sol1}. The $rms$ of the residuals is 2.9\ms \ when this signal is
removed.
The $O-C$ residuals do show some pattern when plotted against the FWHM of the cross-correlation function, meaning that chromospheric activity is still present, but at a lower level that previously thought.

\section{Discussion}
\label{ccl}

RV Doppler signals due to planets are plagued by stellar activity \citep[e.g.,][]{saar98,desort07,boisse09}. Even searching for giant planets, this may remain the main problem. Removing activity signatures \citep[e.g.,][]{dumusque11,haywood14} is only possible when the highest precision is obtained on the RV measurement and on other parameters of the cross-correlation function and when the highest S/N spectra are collected, since the CaII emission is measured on the bluest part of the spectra, which receive the lowest photon count on these relatively cool objects. If activity correction is expected to improve the extraction of the planet signal, activity indicators need to be measured with the highest possible precision. In the sample of measurements discussed in this article, there are at least three cases where stellar activity plays a role: i) the identification of the hot-Jupiter planetary signal of \a, ii) the anterior misinterpretation of the signal of \f, and iii) the power distribution in the periodogram of the CaII measurements for \b. In the first case, strong jitter masked out the main planetary signal; without any clear correlation between the bisector span and the radial velocity and with a small $\log R'_{\mathrm{HK}}$ , it was difficult to assess the real impact of the stellar activity on the RV signal. Even for a $K$ = 89 \ms\ semi-amplitude Doppler wobble, the star required extensive follow-up over several years. The complexity of stellar surfaces in some stars prevents a clear BIS-RV relation from appearing, as in \citet{boisse09, melo}, and an auto-correction of the cross-correlation function (i.e., removal of the BIS-RV slope) is impossible or would just add random noise. In the second case, the RV time series of \f showed a significant correlation with the CaII index and the bisector span based on the first five years of measurements, and these correlations are not significant anymore after five additional years of observations. In the last case, there is a peak near the orbital period in the Lomb-Scargle periodogram of the CaII measurements of \b, although its main properties show a quiet solar-like star. Dispersion in the activity indicators of \b is also seen, and the signal amplitudes are small, which complicates the global interpretation. In all three cases, we are confident about the planetary signals as interpreted in Sect. \ref{planet}, but future measurements could induce modifications of these conclusions.

The long-term evolution of all these activity parameters and their dependencies on the radial velocity are thus difficult to understand, especially with a limited amount of observing time per target in a volume-limited survey, where the highest precision or the densest observing cadence cannot be achieved.

In the case of hot Jupiters, it has been expected that massive planets close to the star surface may induce extra photospheric activity \citep{shkolnik03}. The host star of the hot Jupiter HD 130720 b is an active star, and we searched for the potential signature of activity of such a star in our measurements. With a stellar rotation period of about 10 days, and thus at a synodic period of $\sim$8 days, no visible trend is visible in our data, however, the S/N of the individual CaII measurements is much
lower than in previous similar attempts \citep{shkolnik08,fares12}. As expected, we found no period in the CaII index that could indicate induced activity.
In addition, the $Galex$ flux of \a in the far-UV and near-UV domain was compared with the general population of stars with hot Jupiters and distant Jupiters, as discussed in \citet{shkolnik13}: the UV flux values are similar to those of other planet hosts, and this new case fits well with the conclusions of this study (i.e., no clear impact on the UV stellar flux from the close-in planet).\\

The long-period giant planets constitute the important population from which the overlap with the direct imaging method is possible. Of the new planets presented in this work, the system around \e is the most promising regarding angular separation, as a planet-star separation of 0.08 and 0.15 arcsec is expected for the inner and outer planets. This corresponds to 1.5 and 3 times the resolution in the J band and could be within reach of the new-generation extreme adaptive-optics instruments such as ESO/SPHERE and Gemini/GPI. However, the system is evolved (age of several Gyr), and both these planets have a minimum mass smaller than the Jupiter mass, and their intrinsic luminosity is extremely faint (absolute magnitude 35 or more) while the reflected luminosity is damped by a too large orbital distance.

The new planets have been added to the known population of giant exoplanets (filled circles in Fig. \ref{smam}). \a b piles up with other hot Jupiters of mass lower than Jupiter's, with a period slightly longer than other, more massive giant planets; it is a close twin of $\upsilon$ And b, whose host star is significantly more massive, however (1.27 \msol instead of 0.79 \msol). All other planets are giant and distant planets with minimum masses between those of Saturn and Jupiter and semi-major axes between 1 and 3 AU, hence in populated areas of the mass and orbital distance parameter space, as predicted by the core-accretion formation model \citep[][and references therein]{mordasini09,ida13,benz14}.

From the new sample of giant exoplanets reported in this work, only HD 564 b lies in the habitable zone (HZ) of its parent star. The planet around \a is too hot  and planets around \c,\d, \e, and \f are too cold according to the HZ definition of \citet{kasting93}. \b b has a mass similar to that of Saturn; if it were, as Saturn, orbited by a set of moons, especially large ones like Titan, life as we know it could probably emerge and develop on these moons \citep{williams97}. Detection of such a moon with the current methods and instruments is extremely challenging, however, as outlined by \citet{kipping14}.

Finally, the systems characterized here would strongly benefit from additional observations. As discussed above, such planets in a period range longer than three years with a semi-amplitude RV signal of a few tens of \ms\ are numerous, and the planet parameters may vary with time when the full period of the orbit is not covered, the stellar activity is somewhat high, or/and the observational signal-to-noise ratio is too low for a precise measurement of the activity indicators. A long observing sequence allows either to average the activity signal down, to select parts of the sequence that are less affected, or to reveal additional companions that create the RV jitter. Long-term observational efforts using ELODIE/SOPHIE \citep{boisse12}, Lick/Keck-HIRES  \citep{fischer14}, CORALIE  \citep{marmier13}, and HARPS  \citep{locurto10,mayor11} should thus be followed-up as long as the hardware and facilities exist and perform at a precision level of a few \ms.

\begin{table*}
\caption{Observed and inferred stellar parameters for the planet-hosting stars presented here. For stars with lowest masses, the relationships that derive the activity index  $\log R'_{\mathrm{HK}}$ and the projected velocity are not reliable, so no value is given. }
\label{TableStars}
\centering
\begin{tabular}{l l c c c c c c}
\hline\hline
\multicolumn{2}{l}{\bf Parameter} & \bf \a      & \bf \b                & \bf \c          & \bf \d                & \bf \e                        & \bf \f \\
\hline 
Sp                              &       & K3 V          & G2/G3 V       & G9 V          &  K2 V                 &  K9 V                 & K7 V  \\
$V$ & [mag]                     & 9.49          &  8.29                 & 9.11            & 9.36          & 9.057                 & 9.99\\
$B-V$ & [mag]                   &  0.95         & 0.61                  &  0.82           &  0.93                 &  1.376                        &1.22\\
$\pi$ & [mas]                   & 24.05  (1.20) &  18.64 (0.88) & 17.51 (1.2)   & 20.23 (0.93)  &  63.03 (1.36)         & 36.65 (1.73)\\
$d$ & [pc]                      &  41.6                 & 53.6                 & 57.1          & 49.4          & 15.9                  & 27.3  \\
$T_{\mathrm{eff}}$ & [K]        &5017 (88)      & 5902 (61)     & 5400 (74)       &5122 (79)      &  4462 (145)           & 4475  (100) \\
log $g$            & [cgs]      &  4.43 (0.16)  & 4.53 (0.10)         & 4.37 (0.55)   &4.45 (0.14)    &3.79  (0.53)           & 4.10  (0.36) \\
$\mathrm{[Fe/H]}$  & [dex]& -0.02 (0.06)        & -0.20 (0.05)         &0.13 (0.06)    &0.04 (0.06)    & -0.24 (0.06)          & -0.48   (0.05)\\
$M_*$ & [M$_{\odot}$]   & 0.794 (0.040) &0.961 (0.050) &  0.92 (0.03)  &0.843 (0.024) &  0.585 (0.05)  & 0.571 (0.014) \\
$v\sin{i}$ & [km s$^{-1}$]      & 0.42          & $<$ 2         & 1.7             &  $<$ 2                &  $<$ 2                        &  $<$ 2\\
min/max $S_{MW}$ &      & 0.49/0.62     & 0.12/0.22     & 0.12/0.20     & 0.10/0.27       & 0.81/1.34             & 0.62/0.98 \\
min/max $\log R'_{\mathrm{HK}}$ && -4.53/-4.43& -5.41/-4.68& -5.24/-4.88& -5.30/-4.81         & -                            & -   \\
Age &[Gy]                               &3.8 (3.7)      & 5.5 (3.1)           &      4.8 (4.0)   &  4.9 (4.1)         & 4.3 (4.0)                   & 4.4 (4.0) \\
$R_*$ & [R$_{\odot}$]   &0.73 (0.02)    & 1.01 (0.05)   &   0.91 (0.04) & 0.79 (0.03)    & 0.53 (0.02)           &0.52 (0.02) \\
P$_{rot}$       &[d]                    & 11                    & 10                      & 35                    & 28    &       -       & -\\             
\hline
\end{tabular}
\label{stars}
\end{table*}

\begin{table*} 
\caption{Orbital and physical parameters for five planets presented in
this paper; median values are shown, and 1-$\sigma$ errors are given in parenthesis. $T$ is the epoch of maximum RV. $\sigma$(O-C) is the residual
noise after orbital fitting of the combined set of measurements. }
\label{TablePlanets}
\centering
\begin{tabular}{l l c c c c c c}
\hline\hline
\multicolumn{2}{l}{\bf Parameter}&
& \bf \a\,b & \bf \b\,b & \bf \c\,b & \bf \d\,b  & \bf \f\,b   \\
\hline
$P$ & [days] &                                                  &4.5557 (0.0001)&492.3 (2.3)&1684 (61)  &1129 (-8,+6)&1667 (-31,+33) \\
$T$ & [JD-2454000] &                                    &1387.46 (0.04)&1433 (9)&3181 (-1395,+101)       &2633 (-88,+33)&1899 (-46,38)  \\
  $e$ &            &                                            &0.086 (0.024)&0.096 (0.067)&0.18 (0.15)&0.85 (-0.08,+0.09)&0.05 (-0.04,+0.06) \\
$\gamma$ & [km s$^{-1}$] &                              &-6.5 (1.4)&11.46 (0.38)&     65.8    (0.8)   &56.7 (-4,+3)&-87.3 (0.68) \\ 
$\omega$ & [deg]    &                                   &262 (15)&314 (31)   &82 (-36,+43)   &190 (-5,+2)&231 (-143,+77)\\
$K$ & [m s$^{-1}$] &                                    &89 (2)&8.79 (0.45)  &8.6 (1.1)      &144 (-66,+311)&13.4 (1.0) \\
$a_1 \sin{i}$ & [10$^{-3}$ AU] &                        &0.0371 (0.0008)&0.39 (0.02)&1.30 (0.15)        &2.96 (-0.5,+4.2)& 2.05 (0.16)\\ 
$m_2 \sin{i}$ & [M$_{\mathrm{Jup}}$] &          &0.620 (0.025)&0.33 (0.03)&0.47 (0.06)      &3.5 (-1.2,+3.4)&0.53 (0.05)  \\
$a$ & [AU] &                                                    &0.0498 (0.0008)&1.2 (0.02)&2.69 (0.08) &2.00 (-0.04,+0.04)&2.28 (0.07) \\
\hline                                  
$N_{\mathrm{meas}}$ & &                         &70&99  &46&24&40 \\
$Span$ & [days]       &                                         &3353&4008      &3799&3770&3271 \\
$\sigma$ (O-C) & [m s$^{-1}$] &                         &13.1&2.9&3.6&1.5&2.9\\
$\chi^2_{red}$& &                                               &50 &2.2&6.1&1.1&2.7 \\
\hline
\label{sol1}
\end{tabular}
\end{table*}

\begin{table*} 
\caption{Orbital and physical parameters for the two planets around \e presented in
this paper; {\bf median values are shown, and 1-$\sigma$ errors are given in parenthesis.} }
\label{TablePlanets}
\centering
\begin{tabular}{l l c c c}
\hline\hline
\multicolumn{2}{l}{\bf Parameter}&
& \bf \e\,b & \bf \e\,c \\
\hline
$P$ & [days] &                                                  & 663.2 (-7.4,+8.4)&1818 (-22,+25) \\
$T$ & [JD-2454000] &                                    &1500 (17)&2741 (31) \\
  $e$ &            &                                            &0.14 (0.08)&0.20 (0.04) \\
$\omega$ & [deg]    &                                   &74 (-29,+37)&280 (16) \\
$K$ & [m s$^{-1}$] &                                    &12.2 (1.1)&22.6 (0.8) \\
$a_1 \sin{i}$ & [10$^{-3}$ AU] &                        &0.73 (0.06)&3.70 (0.13) \\ 
$m_2 \sin{i}$ & [M$_{\mathrm{Jup}}$] &          &0.36 (0.04)&0.93 (0.06) \\
$a$ & [AU] &                                                    &1.24 (0.04)&2.44 (0.07) \\
\hline                                  
$\gamma$ & [km s$^{-1}$] &              39.2 (0.6)\\ 
$N_{\mathrm{meas}}$ &           &       75 \\
$Span$ & [days]       &                         3771 \\
$\sigma$ (O-C) & [m s$^{-1}$] & 3.5\\
$\chi^2_{red}$& &                               3.4 \\
\hline
\label{sol2}
\end{tabular}
\end{table*}

\begin{acknowledgements}
NCS and SGS acknowledge the support from Funda\c{c}\~ao para a Ci\^encia e a Tecnologia (FCT, Portugal) through FEDER funds in program COMPETE, as well as through national funds, in the form of grants reference
RECI/FIS-AST/0176/2012 (FCOMP-01-0124-FEDER-027493) and
RECI/FIS-AST/0163/2012 (FCOMP-01-0124-FEDER-027492). 
We also acknowledge the support from the European Research Council/European Community under the FP7 through Starting Grant agreement number 239953. NCS was supported by FCT through the Investigador FCT contract reference IF/00169/2012 and POPH/FSE (EC) by FEDER funding through the program "Programa Operacional de Factores de Competitividade - COMPETE". SGS was supported by FCT through the grant with the reference SFRH/BPD/47611/2008. We are grateful to the ESO staff for their support during observations. \end{acknowledgements}

\bibliographystyle{aa}
\bibliography{references}


\onltab{3}{
\begin{table}[b]
\tiny
\centering
  \caption{Radial-velocity measurements obtained with HARPS of \a. S/N gives the signal-to-noise value per pixel at 550nm.}
  \label{rv1}
\begin{tabular}{lccccc}
\hline
JD-2,400,000.  &  RV & $\sigma_{RV}$ & BIS & S$_{MW}$ & S/N\\
         & [km s$^{-1}$]   &  [km s$^{-1}$] &  [km s$^{-1}$] &  & \\
\hline
53410.725537 & -6.42457 & 0.00158 & 0.01744 & 0.622044 & 53.40 \\
53410.750724 & -6.42760 & 0.00158 & 0.02400 & 0.616814 & 53.30 \\
53412.732968 & -6.61754 & 0.00178 & 0.05596 & 0.614279 & 49.10 \\
53542.554647 & -6.43757 & 0.00276 & 0.03710 & 0.596911 & 33.20 \\
53764.875570 & -6.49020 & 0.00173 & 0.03076 & 0.570907 & 47.70 \\
53765.779982 & -6.42867 & 0.00166 & 0.01368 & 0.585120 & 50.00 \\
54166.758756 & -6.46572 & 0.00159 & 0.02844 & 0.538893 & 51.20 \\
54168.742727 & -6.60277 & 0.00202 & 0.01259 & 0.576631 & 41.80 \\
54170.746496 & -6.44605 & 0.00152 & 0.02771 & 0.575096 & 53.70 \\
54172.764332 & -6.56071 & 0.00152 & 0.02676 & 0.600196 & 54.20 \\
54174.726235 & -6.50707 & 0.00163 & 0.04095 & 0.597889 & 50.50 \\
54229.652114 & -6.46689 & 0.00307 & 0.03563 & 0.602207 & 30.00 \\
54231.664769 & -6.54597 & 0.00256 & 0.05207 & 0.599676 & 34.80 \\
54232.559019 & -6.61496 & 0.00190 & 0.04002 & 0.582196 & 43.80 \\
54234.598066 & -6.44116 & 0.00316 & 0.01298 & 0.548020 & 29.30 \\
54254.493249 & -6.54019 & 0.00265 & 0.05261 & 0.589170 & 33.60 \\
54259.550772 & -6.58628 & 0.00221 & 0.03511 & 0.539361 & 38.20 \\
54568.647269 & -6.53749 & 0.00191 & 0.02419 & 0.504989 & 43.10 \\
54569.640520 & -6.60856 & 0.00138 & 0.02683 & 0.516682 & 56.80 \\
54570.648595 & -6.55564 & 0.00171 & 0.00828 & 0.537879 & 47.00 \\
55325.571974 & -6.58948 & 0.00253 & 0.04524 & 0.535369 & 36.20 \\
55577.838644 & -6.50951 & 0.00207 & 0.01567 & 0.521618 & 42.90 \\
55580.784534 & -6.59489 & 0.00227 & 0.02530 & 0.562490 & 40.80 \\
55586.829408 & -6.51970 & 0.00167 & 0.00917 & 0.537119 & 52.10 \\
55587.842698 & -6.41890 & 0.00173 & 0.01321 & 0.544208 & 50.70 \\
55616.703495 & -6.54845 & 0.00129 & 0.03539 & 0.569496 & 66.90 \\
55637.714149 & -6.46446 & 0.00226 & 0.04577 & 0.525493 & 39.70 \\
55657.704287 & -6.55822 & 0.00175 & 0.04059 & 0.552351 & 50.00 \\
55658.726448 & -6.61764 & 0.00276 & 0.04118 & 0.517552 & 34.30 \\
55659.606019 & -6.55516 & 0.00316 & 0.03756 & 0.524144 & 30.70 \\
55660.662655 & -6.44801 & 0.00167 & 0.02931 & 0.530118 & 51.70 \\
55662.628820 & -6.58349 & 0.00136 & 0.02880 & 0.557618 & 61.70 \\
55663.657096 & -6.61054 & 0.00175 & 0.02668 & 0.566528 & 49.80 \\
55685.596119 & -6.58948 & 0.00165 & 0.02840 & 0.586229 & 52.50 \\
55686.566390 & -6.61100 & 0.00191 & 0.03462 & 0.569396 & 46.10 \\
55690.670990 & -6.61621 & 0.00226 & 0.02865 & 0.533763 & 40.10 \\
55692.639087 & -6.44475 & 0.00244 & 0.04575 & 0.529892 & 37.40 \\
55693.560870 & -6.48774 & 0.00184 & 0.02761 & 0.531775 & 47.40 \\
55695.513106 & -6.61786 & 0.00184 & 0.03668 & 0.525540 & 47.70 \\
55714.590148 & -6.51440 & 0.00151 & 0.03540 & 0.564957 & 56.20 \\
55945.872673 & -6.62038 & 0.00225 & 0.05078 & 0.550842 & 41.20 \\
55947.857270 & -6.45361 & 0.00267 & 0.05497 & 0.524203 & 35.00 \\
55948.841893 & -6.50988 & 0.00176 & 0.03739 & 0.529801 & 50.10 \\
56005.704890 & -6.57270 & 0.00185 & 0.03797 & 0.572106 & 48.10 \\
56006.681664 & -6.45683 & 0.00241 & 0.03736 & 0.561616 & 38.80 \\
56007.675207 & -6.47036 & 0.00291 & 0.03292 & 0.568068 & 33.30 \\
56008.688587 & -6.56441 & 0.00302 & 0.04218 & 0.586245 & 32.20 \\
56009.754245 & -6.61010 & 0.00187 & 0.03584 & 0.568689 & 48.10 \\
56020.644924 & -6.43282 & 0.00260 & 0.02495 & 0.579611 & 35.90 \\
56051.689679 & -6.51262 & 0.00248 & 0.02083 & 0.566476 & 37.90 \\
56052.530097 & -6.43495 & 0.00266 & 0.02445 & 0.565486 & 35.30 \\
56053.501755 & -6.48342 & 0.00253 & 0.03189 & 0.538947 & 37.40 \\
56054.588635 & -6.59068 & 0.00386 & 0.04458 & 0.497023 & 26.20 \\
56055.573461 & -6.59932 & 0.00242 & 0.04659 & 0.545490 & 37.80 \\
56056.626337 & -6.45720 & 0.00222 & 0.03269 & 0.547406 & 40.70 \\
56057.485811 & -6.42548 & 0.00198 & 0.02057 & 0.549510 & 46.20 \\
56058.524010 & -6.51691 & 0.00261 & 0.02659 & 0.551157 & 35.80 \\
56059.607215 & -6.60978 & 0.00181 & 0.03832 & 0.553824 & 49.40 \\
56060.575567 & -6.54854 & 0.00355 & 0.03068 & 0.540806 & 28.20 \\
56061.589262 & -6.43499 & 0.00256 & 0.02865 & 0.609882 & 36.30 \\
56062.561883 & -6.46946 & 0.00236 & 0.02492 & 0.618299 & 38.70 \\
56080.511699 & -6.44618 & 0.00170 & 0.03021 & 0.583868 & 52.60 \\
56318.829398 & -6.58097 & 0.00212 & 0.02528 & 0.618238 & 44.70 \\
56319.790147 & -6.61293 & 0.00312 & 0.04747 & 0.590154 & 32.00 \\
56365.732344 & -6.54633 & 0.00203 & 0.02942 & 0.572228 & 44.40 \\
56385.740518 & -6.46192 & 0.00253 & 0.04090 & 0.560083 & 38.10 \\
56696.877943 & -6.60508 & 0.00130 & 0.03711 & 0.512218 & 65.50 \\
56723.694284 & -6.53963 & 0.00192 & 0.03643 & 0.534451 & 46.30 \\
56727.758738 & -6.50879 & 0.00220 & 0.03820 & 0.531676 & 40.80 \\
56738.742879 & -6.59142 & 0.00143 & 0.01051 & 0.541591 & 59.70 \\
56763.650168 & -6.47696 & 0.00123 & 0.05046 & 0.536203 & 67.70 \\
\hline
\end{tabular}
\end{table}
}

\onltab{4}{
\begin{table}[b]
\tiny
\centering
  \caption{Radial-velocity measurements obtained with HARPS of \b. S/N gives the signal-to-noise value per pixel at 550nm.}
  \label{rv2}
\begin{tabular}{lccccc}
\hline
JD-2,400,000.  &  RV & $\sigma_{RV}$ & BIS & S$_{MW}$ & S/N \\
         & [km s$^{-1}$]   &  [km s$^{-1}$] &  [km s$^{-1}$] &  & \\
\hline
52851.876768 & 11.46301 & 0.00116 & -0.02048 & 0.184060 & 126.30 \\
52853.869168 & 11.46523 & 0.00284 & -0.02295 & 0.182541 & 43.10 \\
52941.607800 & 11.48328 & 0.00566 & -0.00832 & 0.162482 & 23.40 \\
53264.713004 & 11.44450 & 0.00220 & -0.00873 & 0.180716 & 53.50 \\
53266.724775 & 11.45493 & 0.00197 & -0.01473 & 0.184430 & 55.80 \\
53271.732825 & 11.45242 & 0.00414 & -0.01653 & 0.185826 & 30.90 \\
53274.738912 & 11.45120 & 0.00284 & -0.03438 & 0.143082 & 41.00 \\
53288.679402 & 11.45507 & 0.00259 & -0.00705 & 0.191112 & 40.70 \\
53291.680278 & 11.45266 & 0.00426 & -0.02654 & 0.184789 & 28.20 \\
53294.677291 & 11.46005 & 0.00286 & -0.01219 & 0.178598 & 35.10 \\
53296.692360 & 11.45192 & 0.00290 & -0.01498 & 0.188696 & 36.10 \\
53298.715144 & 11.45380 & 0.00275 & -0.01561 & 0.178162 & 36.40 \\
53310.662439 & 11.46287 & 0.00227 & -0.02000 & 0.183368 & 48.60 \\
53721.566682 & 11.45269 & 0.00160 & -0.01758 & 0.161832 & 61.40 \\
53870.933543 & 11.45767 & 0.00226 & -0.02449 & 0.174302 & 44.60 \\
53944.861248 & 11.47027 & 0.00142 & -0.01619 & 0.173747 & 68.70 \\
53948.860547 & 11.46677 & 0.00456 & -0.02274 & 0.210435 & 25.10 \\
53980.770152 & 11.46982 & 0.00166 & -0.02086 & 0.170453 & 58.70 \\
54384.726844 & 11.46376 & 0.00231 & -0.01591 & 0.197163 & 44.10 \\
54385.634255 & 11.45947 & 0.00529 & 0.00065 & 0.211525 & 22.50 \\
54391.641142 & 11.47579 & 0.00223 & -0.01759 & 0.188990 & 43.90 \\
54393.663465 & 11.47829 & 0.00291 & -0.01915 & 0.195049 & 34.90 \\
54395.681422 & 11.47337 & 0.00216 & -0.01004 & 0.190786 & 45.30 \\
54420.585952 & 11.47179 & 0.00185 & -0.01240 & 0.177270 & 53.70 \\
54423.589368 & 11.47090 & 0.00298 & -0.02655 & 0.171492 & 35.30 \\
54424.516514 & 11.47099 & 0.00259 & -0.01498 & 0.170450 & 39.70 \\
54426.542184 & 11.46958 & 0.00212 & -0.01332 & 0.174664 & 46.10 \\
54429.520725 & 11.47675 & 0.00226 & -0.01690 & 0.178376 & 44.80 \\
54430.643552 & 11.47348 & 0.00217 & -0.00422 & 0.177047 & 46.70 \\
54437.595286 & 11.47218 & 0.00157 & -0.01269 & 0.180847 & 62.90 \\
54446.594441 & 11.47554 & 0.00231 & -0.01918 & 0.176365 & 44.60 \\
54448.541103 & 11.47121 & 0.00272 & -0.02339 & 0.182717 & 38.30 \\
54450.542773 & 11.47128 & 0.00266 & -0.02895 & 0.203111 & 38.90 \\
54609.908159 & 11.46230 & 0.00223 & -0.01515 & 0.179853 & 45.70 \\
54617.933002 & 11.45791 & 0.00249 & -0.02531 & 0.200989 & 41.60 \\
54637.902968 & 11.45669 & 0.00200 & -0.01659 & 0.178432 & 50.80 \\
54670.781602 & 11.45625 & 0.00240 & -0.02241 & 0.157861 & 43.20 \\
54675.855760 & 11.45714 & 0.00179 & -0.01883 & 0.176921 & 54.90 \\
54687.930372 & 11.45068 & 0.00750 & -0.02029 & 0.136157 & 17.90 \\
54698.797546 & 11.46021 & 0.00213 & -0.02191 & 0.196208 & 46.40 \\
54709.844063 & 11.45673 & 0.00152 & -0.01328 & 0.184255 & 65.60 \\
54719.755513 & 11.45500 & 0.00266 & -0.01147 & 0.196230 & 39.30 \\
54722.825809 & 11.45542 & 0.00153 & -0.01793 & 0.181264 & 64.60 \\
54730.740618 & 11.45325 & 0.00318 & -0.02554 & 0.192506 & 34.30 \\
54751.725433 & 11.45327 & 0.00144 & -0.01583 & 0.188489 & 68.70 \\
54752.622288 & 11.45613 & 0.00188 & -0.01105 & 0.193610 & 52.30 \\
54778.571503 & 11.46036 & 0.00145 & -0.01336 & 0.181394 & 67.30 \\
54812.570368 & 11.46080 & 0.00304 & -0.02692 & 0.158688 & 34.80 \\
55016.915860 & 11.46173 & 0.00271 & -0.01891 & 0.161838 & 36.90 \\
55018.861671 & 11.46745 & 0.00183 & -0.01943 & 0.172494 & 54.70 \\
55079.807536 & 11.45950 & 0.00136 & -0.02195 & 0.176188 & 72.30 \\
55090.789049 & 11.46052 & 0.00146 & -0.02214 & 0.170126 & 67.90 \\
55091.643857 & 11.45106 & 0.00481 & -0.03331 & 0.116738 & 24.40 \\
55105.694107 & 11.45433 & 0.00328 & -0.01311 & 0.146626 & 33.30 \\
55107.683151 & 11.45683 & 0.00394 & -0.02926 & 0.165819 & 28.60 \\
55136.581223 & 11.45795 & 0.00770 & -0.01947 & 0.221511 & 17.20 \\
55217.538712 & 11.45284 & 0.00253 & -0.01375 & 0.179174 & 41.80 \\
55349.887745 & 11.46477 & 0.00175 & -0.02466 & 0.189634 & 59.30 \\
55352.952710 & 11.46457 & 0.00211 & -0.00762 & 0.165068 & 50.60 \\
55355.889795 & 11.46512 & 0.00258 & -0.02188 & 0.158900 & 41.50 \\
55388.885020 & 11.46253 & 0.00212 & -0.01438 & 0.176285 & 49.20 \\
55391.894404 & 11.47003 & 0.00220 & -0.03096 & 0.167251 & 48.90 \\
55397.867729 & 11.47196 & 0.00424 & -0.01453 & 0.137993 & 27.60 \\
55401.791292 & 11.47254 & 0.00509 & -0.00882 & 0.151250 & 23.80 \\
55404.844242 & 11.46508 & 0.00227 & -0.02066 & 0.167329 & 45.50 \\
55405.930310 & 11.47157 & 0.00164 & -0.01824 & 0.154994 & 64.70 \\
55409.891367 & 11.47203 & 0.00171 & -0.01425 & 0.173069 & 62.70 \\
55424.709058 & 11.46882 & 0.00402 & -0.02515 & 0.177195 & 29.00 \\
55433.869077 & 11.47036 & 0.00154 & -0.01876 & 0.167255 & 68.90 \\
55434.713976 & 11.47563 & 0.00232 & -0.01409 & 0.162711 & 46.10 \\
55445.650274 & 11.47416 & 0.00228 & -0.01857 & 0.169120 & 47.10 \\
55453.679814 & 11.46923 & 0.00194 & -0.02907 & 0.174891 & 54.10 \\
55457.614894 & 11.47464 & 0.00199 & -0.03012 & 0.158611 & 53.00 \\
55460.786864 & 11.47464 & 0.00148 & -0.01968 & 0.175336 & 70.00 \\
55482.664228 & 11.46709 & 0.00102 & -0.02422 & 0.174177 & 101.00 \\
55490.774792 & 11.46720 & 0.00205 & -0.01445 & 0.156100 & 52.70 \\
55520.603912 & 11.46248 & 0.00336 & -0.01886 & 0.139505 & 33.00 \\
55770.909913 & 11.46034 & 0.00159 & -0.01441 & 0.181841 & 65.70 \\
55782.894474 & 11.45780 & 0.00187 & -0.01196 & 0.179220 & 54.60 \\
56115.896670 & 11.45713 & 0.00116 & -0.01535 & 0.188004 & 90.80 \\
56150.766066 & 11.45566 & 0.00349 & -0.02903 & 0.168490 & 33.20 \\
56169.766216 & 11.45456 & 0.00223 & -0.01079 & 0.192058 & 48.30 \\
56173.851140 & 11.45544 & 0.00170 & -0.01593 & 0.194119 & 63.80 \\
56174.794849 & 11.46014 & 0.00311 & -0.01688 & 0.202797 & 36.80 \\
56175.913555 & 11.45429 & 0.00365 & -0.01435 & 0.182058 & 32.90 \\
56207.734956 & 11.45833 & 0.00217 & -0.01906 & 0.185766 & 49.80 \\
56239.619268 & 11.45269 & 0.00170 & -0.01765 & 0.171106 & 62.70 \\
56253.606856 & 11.45868 & 0.00153 & -0.01731 & 0.175562 & 71.10 \\
56269.598885 & 11.46143 & 0.00191 & -0.02148 & 0.180665 & 56.90 \\
56271.533034 & 11.45986 & 0.00142 & -0.01874 & 0.187842 & 78.80 \\
56289.545712 & 11.46245 & 0.00135 & -0.01809 & 0.176270 & 80.50 \\
56304.523400 & 11.46253 & 0.00172 & -0.02680 & 0.167858 & 64.70 \\
56506.912689 & 11.46897 & 0.00212 & -0.02075 & 0.164968 & 51.20 \\
56532.870914 & 11.46497 & 0.00309 & -0.01780 & 0.158232 & 37.20 \\
56555.807983 & 11.46251 & 0.00108 & -0.02263 & 0.168914 & 97.40 \\
56556.866358 & 11.46072 & 0.00124 & -0.01860 & 0.179422 & 85.30 \\
56615.616037 & 11.45646 & 0.00154 & -0.01822 & 0.174320 & 69.30 \\
56620.518959 & 11.45092 & 0.00223 & -0.02245 & 0.165451 & 48.00 \\
\hline
\end{tabular}
\end{table}
}

\onltab{5}{
\begin{table}[b]
\tiny
\centering
  \caption{Radial-velocity measurements obtained with HARPS of \c. S/N gives the signal-to-noise value per pixel at 550nm.}
  \label{rv3}
\begin{tabular}{lccccc}
\hline
JD-2,400,000.  &  RV & $\sigma_{RV}$ & BIS & S$_{MW}$ & S/N \\
         & [km s$^{-1}$]   &  [km s$^{-1}$] &  [km s$^{-1}$] &  & \\
\hline
52946.812523 & 65.86486 & 0.00296 & -0.04371 & 0.169431 & 31.20 \\
53056.566167 & 65.87576 & 0.00195 & -0.04117 & 0.144889 & 45.50 \\
53369.720161 & 65.87603 & 0.00116 & -0.04906 & 0.160031 & 65.80 \\
54431.697383 & 65.86837 & 0.00209 & -0.04986 & 0.149176 & 38.10 \\
54478.695858 & 65.86445 & 0.00137 & -0.04016 & 0.164764 & 56.90 \\
54487.585050 & 65.86567 & 0.00179 & -0.05042 & 0.156202 & 44.00 \\
54720.903534 & 65.87229 & 0.00148 & -0.04588 & 0.169663 & 50.90 \\
54736.855063 & 65.87009 & 0.00185 & -0.04634 & 0.173602 & 41.90 \\
54814.693440 & 65.87700 & 0.00162 & -0.04665 & 0.150827 & 46.90 \\
54841.635439 & 65.87511 & 0.00142 & -0.03868 & 0.168755 & 52.20 \\
54872.604317 & 65.87586 & 0.00202 & -0.04659 & 0.151099 & 39.90 \\
54902.527097 & 65.87319 & 0.00133 & -0.04828 & 0.165943 & 57.30 \\
54902.533138 & 65.87281 & 0.00135 & -0.03476 & 0.165018 & 56.70 \\
55105.808888 & 65.87827 & 0.00380 & -0.04197 & 0.126462 & 24.00 \\
55109.822612 & 65.87898 & 0.00314 & -0.03762 & 0.149445 & 27.20 \\
55218.690341 & 65.88129 & 0.00255 & -0.05587 & 0.168758 & 33.80 \\
55260.550527 & 65.88385 & 0.00132 & -0.04528 & 0.158792 & 58.90 \\
55264.532568 & 65.88679 & 0.00131 & -0.04521 & 0.164005 & 59.20 \\
55434.908336 & 65.89159 & 0.00179 & -0.04079 & 0.139469 & 45.20 \\
55490.873740 & 65.88203 & 0.00179 & -0.04018 & 0.142034 & 44.10 \\
55587.538203 & 65.89119 & 0.00191 & -0.04806 & 0.129387 & 43.20 \\
55653.511919 & 65.88355 & 0.00167 & -0.04979 & 0.150054 & 50.50 \\
55903.735078 & 65.88064 & 0.00210 & -0.04806 & 0.134130 & 39.80 \\
55962.518684 & 65.87736 & 0.00186 & -0.04236 & 0.145218 & 44.80 \\
55966.562222 & 65.87388 & 0.00106 & -0.04491 & 0.147531 & 77.80 \\
55968.629501 & 65.87349 & 0.00118 & -0.03987 & 0.150913 & 70.20 \\
56149.914280 & 65.86335 & 0.00125 & -0.04473 & 0.147855 & 63.80 \\
56160.905614 & 65.87109 & 0.00265 & -0.04541 & 0.127865 & 33.50 \\
56167.891515 & 65.87172 & 0.00206 & -0.04419 & 0.148821 & 40.30 \\
56168.878684 & 65.87151 & 0.00177 & -0.03894 & 0.133997 & 46.60 \\
56171.855514 & 65.87650 & 0.00120 & -0.04169 & 0.160186 & 67.80 \\
56173.912158 & 65.87852 & 0.00145 & -0.04756 & 0.178192 & 56.20 \\
56181.837185 & 65.87365 & 0.00135 & -0.04721 & 0.146710 & 60.10 \\
56191.919832 & 65.87574 & 0.00127 & -0.04719 & 0.152157 & 65.30 \\
56204.832233 & 65.87369 & 0.00167 & -0.04065 & 0.181777 & 49.00 \\
56208.878254 & 65.86995 & 0.00215 & -0.04075 & 0.189935 & 38.50 \\
56209.872861 & 65.86819 & 0.00315 & -0.04039 & 0.199631 & 29.00 \\
56251.698978 & 65.87025 & 0.00132 & -0.04441 & 0.145983 & 63.90 \\
56256.788837 & 65.86819 & 0.00104 & -0.04280 & 0.153352 & 82.90 \\
56289.697046 & 65.87007 & 0.00129 & -0.04239 & 0.145702 & 62.30 \\
56356.527585 & 65.87257 & 0.00182 & -0.04367 & 0.173913 & 48.00 \\
56391.492430 & 65.87434 & 0.00198 & -0.03824 & 0.131935 & 45.10 \\
56543.870683 & 65.87354 & 0.00193 & -0.05104 & 0.141112 & 43.90 \\
56564.828467 & 65.87315 & 0.00194 & -0.04800 & 0.145339 & 43.30 \\
56615.788648 & 65.88156 & 0.00111 & -0.04352 & 0.161555 & 71.40 \\
56722.552843 & 65.88013 & 0.00168 & -0.04489 & 0.141415 & 49.60 \\
56745.477862 & 65.87729 & 0.00175 & -0.03852 & 0.164149 & 49.10 \\
56745.481647 & 65.87466 & 0.00157 & -0.03779 & 0.159891 & 53.60 \\
\hline
\end{tabular}
\end{table}
}

\onltab{6}{
\begin{table}[b]
\tiny
\centering
  \caption{Radial-velocity measurements obtained with HARPS of \d.  S/N gives the signal-to-noise value per pixel at 550nm.}
  \label{rv5}
\begin{tabular}{lccccc}
\hline
JD-2,400,000.  &  RV & $\sigma_{RV}$ & BIS & S$_{MW}$ & S/N \\
         & [km s$^{-1}$]   &  [km s$^{-1}$] &  [km s$^{-1}$] &  & \\
\hline
53047.806558 & 56.73236 & 0.00191 & -0.00523 & 0.222966 & 46.40 \\
53056.826970 & 56.73343 & 0.00158 & 0.00088 & 0.206144 & 56.00 \\
53515.599740 & 56.72315 & 0.00190 & -0.00463 & 0.157178 & 41.30 \\
54921.798313 & 56.69606 & 0.00144 & 0.00549 & 0.226998 & 51.40 \\
55338.656931 & 56.72967 & 0.00372 & 0.01091 & 0.267985 & 25.00 \\
55674.563476 & 56.72937 & 0.00241 & -0.00064 & 0.170013 & 34.00 \\
55948.863549 & 56.71362 & 0.00166 & 0.00232 & 0.173786 & 48.00 \\
56023.742177 & 56.70398 & 0.00169 & 0.00073 & 0.168008 & 47.10 \\
56025.688336 & 56.70462 & 0.00263 & 0.00176 & 0.166300 & 32.40 \\
56051.736647 & 56.70056 & 0.00130 & 0.00392 & 0.187639 & 58.60 \\
56078.570007 & 56.69199 & 0.00175 & -0.00161 & 0.145563 & 47.20 \\
56094.601475 & 56.68802 & 0.00192 & 0.00986 & 0.164142 & 42.90 \\
56168.472537 & 56.60249 & 0.00216 & 0.00177 & 0.158093 & 39.20 \\
56311.801063 & 56.71854 & 0.00177 & 0.00414 & 0.209571 & 47.10 \\
56332.784952 & 56.72480 & 0.00390 & 0.01071 & 0.205499 & 23.80 \\
56332.806377 & 56.72290 & 0.00232 & 0.00074 & 0.247088 & 36.30 \\
56369.746228 & 56.72588 & 0.00197 & 0.00129 & 0.246852 & 43.30 \\
56409.658101 & 56.72661 & 0.00343 & -0.00126 & 0.113681 & 26.60 \\
56679.870605 & 56.73221 & 0.00260 & -0.00068 & 0.135687 & 33.50 \\
56716.862191 & 56.72798 & 0.00157 & 0.00195 & 0.185636 & 51.40 \\
56731.662512 & 56.73342 & 0.00132 & 0.00499 & 0.185475 & 60.40 \\
56736.840769 & 56.73386 & 0.00154 & 0.00150 & 0.184786 & 51.70 \\
56763.585806 & 56.72894 & 0.00087 & 0.00267 & 0.184067 & 88.30 \\
\hline
\end{tabular}
\end{table}
}

\onltab{7}{
\begin{table}[b]
\tiny
\centering
  \caption{Radial-velocity measurements obtained with HARPS of \e. S/N gives the signal-to-noise value per pixel at 550nm.}
  \label{rv6}
\begin{tabular}{lccccc}
\hline
JD-2,400,000.  &  RV & $\sigma_{RV}$ & BIS & S$_{MW}$ & S/N \\
         & [km s$^{-1}$]   &  [km s$^{-1}$] &  [km s$^{-1}$] &  & \\
\hline
53047.881823 & 39.23173 & 0.00305 & 0.06196 & 1.132572 & 35.90 \\
53054.853357 & 39.23526 & 0.00218 & 0.04908 & 1.081999 & 48.10 \\
54169.797044 & 39.21023 & 0.00163 & 0.04642 & 1.106171 & 54.90 \\
54171.841675 & 39.21269 & 0.00283 & 0.03670 & 1.091107 & 35.40 \\
54231.707362 & 39.20796 & 0.00387 & 0.04717 & 1.024483 & 26.90 \\
54254.558935 & 39.20832 & 0.00247 & 0.03249 & 1.065047 & 38.00 \\
54255.598697 & 39.20774 & 0.00204 & 0.03988 & 1.088968 & 44.80 \\
54257.602953 & 39.21209 & 0.00214 & 0.04304 & 1.140604 & 42.50 \\
54258.605074 & 39.20594 & 0.00182 & 0.03554 & 1.139446 & 48.90 \\
54527.870224 & 39.20544 & 0.00206 & 0.03445 & 1.035098 & 44.80 \\
54558.763909 & 39.21444 & 0.00208 & 0.05435 & 1.086478 & 44.00 \\
54582.711795 & 39.21463 & 0.00152 & 0.04262 & 1.093295 & 57.80 \\
54612.681536 & 39.21956 & 0.00235 & 0.04062 & 1.043033 & 40.20 \\
54670.539688 & 39.23551 & 0.00196 & 0.04093 & 1.008780 & 47.30 \\
54687.513837 & 39.22915 & 0.00248 & 0.03922 & 1.123480 & 39.30 \\
54872.814746 & 39.24546 & 0.00190 & 0.04367 & 1.075356 & 47.40 \\
54930.706596 & 39.25341 & 0.00161 & 0.03461 & 1.080127 & 55.00 \\
54930.791436 & 39.25556 & 0.00162 & 0.04109 & 1.114953 & 55.00 \\
54931.701621 & 39.25591 & 0.00190 & 0.04474 & 1.272736 & 47.80 \\
55013.555076 & 39.23016 & 0.00191 & 0.04243 & 0.964742 & 48.20 \\
55017.548011 & 39.23467 & 0.00111 & 0.03545 & 1.015099 & 80.40 \\
55019.520542 & 39.22898 & 0.00239 & 0.03698 & 1.044004 & 40.30 \\
55229.818454 & 39.22747 & 0.00147 & 0.04034 & 0.936748 & 60.10 \\
55266.818669 & 39.22445 & 0.00157 & 0.03614 & 0.975268 & 56.50 \\
55290.775088 & 39.22982 & 0.00143 & 0.04935 & 1.081048 & 61.90 \\
55303.643575 & 39.22369 & 0.00289 & 0.02272 & 0.936411 & 34.10 \\
55311.831470 & 39.23308 & 0.00177 & 0.03440 & 1.046885 & 53.60 \\
55313.584639 & 39.23289 & 0.00257 & 0.03924 & 1.007605 & 37.90 \\
55315.777596 & 39.23394 & 0.00284 & 0.04650 & 0.815662 & 35.70 \\
55321.615314 & 39.22823 & 0.00246 & 0.03302 & 1.053027 & 39.40 \\
55326.594991 & 39.23243 & 0.00222 & 0.04859 & 1.101968 & 43.10 \\
55334.643626 & 39.22892 & 0.00338 & 0.05334 & 1.024856 & 31.10 \\
55338.667081 & 39.23282 & 0.00291 & 0.04506 & 1.007748 & 34.80 \\
55350.643184 & 39.23347 & 0.00241 & 0.03681 & 1.051212 & 40.30 \\
55358.657970 & 39.23032 & 0.00252 & 0.05919 & 1.077452 & 39.40 \\
55358.668097 & 39.22599 & 0.00287 & 0.04342 & 0.925699 & 35.80 \\
55375.502804 & 39.23396 & 0.00169 & 0.04900 & 1.061762 & 54.70 \\
55402.518558 & 39.23080 & 0.00561 & 0.00931 & 0.931513 & 22.20 \\
55408.526208 & 39.23014 & 0.00304 & 0.04389 & 0.992615 & 34.00 \\
55414.476222 & 39.23212 & 0.00201 & 0.04906 & 0.961157 & 48.00 \\
55425.469477 & 39.23312 & 0.00157 & 0.03978 & 0.920064 & 59.30 \\
55437.485285 & 39.23088 & 0.00302 & 0.03167 & 0.811507 & 34.90 \\
55443.485145 & 39.22965 & 0.00561 & 0.04462 & 1.042132 & 22.30 \\
55586.874738 & 39.22011 & 0.00169 & 0.03624 & 1.092483 & 55.00 \\
55672.667929 & 39.20730 & 0.00207 & 0.03167 & 1.085615 & 45.70 \\
55689.633239 & 39.20769 & 0.00155 & 0.05545 & 1.191228 & 59.10 \\
55691.763736 & 39.20173 & 0.00232 & 0.03523 & 1.082861 & 42.60 \\
55694.553894 & 39.20315 & 0.00189 & 0.03102 & 1.050578 & 49.50 \\
55728.518219 & 39.20413 & 0.00243 & 0.05858 & 1.186629 & 41.10 \\
55729.518201 & 39.20088 & 0.00201 & 0.05703 & 1.138445 & 47.30 \\
55749.499966 & 39.20142 & 0.00218 & 0.04478 & 1.176564 & 44.50 \\
55755.453622 & 39.19367 & 0.00310 & 0.02074 & 1.000829 & 34.00 \\
55772.474746 & 39.20061 & 0.01074 & 0.03613 & 1.138059 & 14.00 \\
55961.887747 & 39.19702 & 0.00181 & 0.04792 & 1.092375 & 52.60 \\
55964.807747 & 39.19658 & 0.00264 & 0.02703 & 1.069307 & 38.50 \\
55967.883196 & 39.19852 & 0.00130 & 0.04082 & 0.957865 & 70.70 \\
56023.756120 & 39.20292 & 0.00249 & 0.03843 & 1.141698 & 39.50 \\
56051.779397 & 39.20250 & 0.00155 & 0.03471 & 1.113070 & 59.20 \\
56054.529353 & 39.20143 & 0.00542 & 0.03528 & 1.129481 & 22.20 \\
56095.540186 & 39.20893 & 0.00313 & 0.04351 & 1.196871 & 33.60 \\
56168.478323 & 39.21476 & 0.00239 & 0.03947 & 1.225962 & 42.10 \\
56311.829678 & 39.21198 & 0.00165 & 0.05030 & 1.201069 & 58.70 \\
56312.850414 & 39.22693 & 0.02383 & 0.05976 & -0.069866 & 7.50 \\
56320.839755 & 39.21498 & 0.00222 & 0.04133 & 1.141358 & 44.90 \\
56331.875251 & 39.20866 & 0.00177 & 0.05142 & 1.218641 & 55.50 \\
56369.743652 & 39.20569 & 0.00231 & 0.04073 & 1.275527 & 43.60 \\
56384.797775 & 39.20373 & 0.00231 & 0.04280 & 1.101498 & 43.10 \\
56412.732946 & 39.21537 & 0.00141 & 0.04831 & 1.215775 & 67.00 \\
56441.473126 & 39.21324 & 0.00305 & 0.05314 & 1.228108 & 34.50 \\
56679.874554 & 39.25385 & 0.00304 & 0.05325 & 1.276441 & 34.80 \\
56716.890700 & 39.25921 & 0.00231 & 0.03585 & 1.285259 & 43.50 \\
56730.869538 & 39.25383 & 0.00216 & 0.03982 & 1.172956 & 45.40 \\
56740.776614 & 39.25155 & 0.00114 & 0.04124 & 1.346276 & 82.30 \\
56746.765923 & 39.25945 & 0.00232 & 0.03245 & 1.318461 & 43.30 \\
56755.780837 & 39.25554 & 0.00233 & 0.04692 & 1.246124 & 42.50 \\
\hline
\end{tabular}
\end{table}
}

\onltab{8}{
\begin{table}[b]
\tiny
\centering
  \caption{Radial-velocity measurements obtained with HARPS of \f. S/N gives the signal-to-noise value per pixel at 550nm.}
  \label{rv7}
\begin{tabular}{lccccc}
\hline
JD-2,400,000.  &  RV & $\sigma_{RV}$ & BIS & S$_{MW}$ & S/N \\
         & [km s$^{-1}$]   &  [km s$^{-1}$] &  [km s$^{-1}$] &  & \\
\hline
53577.638500 & -87.34721 & 0.00236 & 0.04557 & 0.789306 & 39.30 \\
53579.641564 & -87.33761 & 0.00207 & 0.02781 & 0.860043 & 43.80 \\
53862.818306 & -87.33258 & 0.00207 & 0.04196 & 0.726725 & 44.10 \\
53864.809683 & -87.32871 & 0.00179 & 0.03201 & 0.792391 & 49.80 \\
53870.791313 & -87.33006 & 0.00133 & 0.03916 & 0.741974 & 65.50 \\
54174.902721 & -87.32065 & 0.00183 & 0.02689 & 0.785526 & 49.90 \\
54392.497515 & -87.31897 & 0.00192 & 0.03391 & 0.731616 & 47.30 \\
54556.888627 & -87.32936 & 0.00151 & 0.03917 & 0.682344 & 57.60 \\
54582.832192 & -87.32881 & 0.00168 & 0.02971 & 0.713436 & 52.50 \\
54609.771934 & -87.32864 & 0.00137 & 0.03030 & 0.719046 & 63.50 \\
54609.838848 & -87.33008 & 0.00166 & 0.03099 & 0.726367 & 54.10 \\
54671.713924 & -87.33226 & 0.00178 & 0.03223 & 0.639844 & 50.90 \\
54686.586739 & -87.33309 & 0.00135 & 0.03167 & 0.676599 & 65.20 \\
54720.559621 & -87.33249 & 0.00173 & 0.03116 & 0.715382 & 51.50 \\
54722.595695 & -87.33276 & 0.00169 & 0.03792 & 0.694008 & 53.50 \\
54931.904537 & -87.34475 & 0.00162 & 0.03805 & 0.704232 & 54.50 \\
55013.714941 & -87.34569 & 0.00184 & 0.04073 & 0.715019 & 49.30 \\
55091.538435 & -87.33488 & 0.00506 & 0.03153 & 0.851032 & 22.40 \\
55260.902428 & -87.34112 & 0.00146 & 0.04076 & 0.846135 & 61.50 \\
55313.832975 & -87.33405 & 0.00297 & 0.03782 & 0.774970 & 34.20 \\
55337.901286 & -87.34011 & 0.00339 & 0.03820 & 0.766106 & 31.70 \\
55358.865074 & -87.33887 & 0.00197 & 0.04183 & 0.770427 & 49.20 \\
55375.722713 & -87.34447 & 0.00278 & 0.02731 & 0.767840 & 36.30 \\
55388.651469 & -87.33068 & 0.00204 & 0.02902 & 0.775941 & 47.30 \\
55404.661284 & -87.34487 & 0.00286 & 0.03101 & 0.790423 & 35.70 \\
55408.734655 & -87.33044 & 0.00294 & 0.02235 & 0.786248 & 35.20 \\
55438.630840 & -87.33190 & 0.00175 & 0.03596 & 0.717285 & 54.00 \\
55491.507904 & -87.32843 & 0.00208 & 0.03192 & 0.779473 & 46.60 \\
55685.943023 & -87.31756 & 0.00193 & 0.04043 & 0.802957 & 50.50 \\
56023.850615 & -87.31882 & 0.00169 & 0.03319 & 1.012264 & 55.90 \\
56023.890306 & -87.32552 & 0.00222 & 0.04285 & 0.919467 & 44.00 \\
56054.757351 & -87.31429 & 0.00569 & 0.04935 & 0.808981 & 21.50 \\
56079.794608 & -87.31693 & 0.00196 & 0.03802 & 0.772068 & 49.30 \\
56388.921992 & -87.33923 & 0.00318 & 0.04078 & 0.746579 & 33.00 \\
56453.797518 & -87.34113 & 0.00559 & 0.03034 & 0.658049 & 21.80 \\
56457.785445 & -87.33738 & 0.00329 & 0.04085 & 0.782907 & 32.30 \\
56474.667467 & -87.33169 & 0.00448 & 0.05626 & 0.680633 & 25.40 \\
56497.755121 & -87.33078 & 0.00253 & 0.02928 & 0.804986 & 39.80 \\
56564.479769 & -87.34215 & 0.00190 & 0.02707 & 0.618980 & 51.00 \\
56749.896171 & -87.34789 & 0.00270 & 0.04503 & 0.731619 & 37.30 \\
\hline
\end{tabular}
\end{table}
}

\end{document}